\documentclass[12pt]{article}
\usepackage[totalwidth=460truept,totalheight=620truept]{geometry}
\usepackage{latexsym,graphicx,amsfonts,amssymb,amsmath,xcolor,cancel}
\usepackage[hypertexnames=false,hidelinks]{hyperref}

\def\theequation{\arabic{section}.\arabic{equation}}

\renewcommand{\theequation}{\thesection.\arabic{equation}}
\global\arraycolsep=1truept
\renewcommand{\theequation}{\arabic{section}.\arabic{equation}}

\begin{document}

\null

\vskip 1truecm

\begin{center}
{\huge \textbf{A New Quantization Principle}}

\vskip.5truecm

{\huge \textbf{from a Minimally}}

\vskip.5truecm

{\huge \textbf{non Time-Ordered Product}}

\vskip1truecm

\textsl{Damiano Anselmi}

\vskip.5truecm

{\small \textit{Dipartimento di Fisica \textquotedblleft
E.Fermi\textquotedblright , Universit\`{a} di Pisa, Largo B.Pontecorvo 3,
56127 Pisa, Italy}}

{\small \textit{INFN, Sezione di Pisa, Largo B. Pontecorvo 3, 56127 Pisa,
Italy}}

{\small damiano.anselmi@unipi.it}

\vskip1truecm

\textbf{Abstract}
\end{center}

We formulate a new quantization principle for perturbative quantum field
theory, based on a minimally non time-ordered product, and show that it
gives the theories of physical particles and purely virtual particles. Given
a classical Lagrangian, the quantization proceeds as usual, guided\ by the
time-ordered product, up to the common scattering matrix $S$, which
satisfies a unitarity or a pseudounitarity equation. The physical scattering
matrix $S_{\text{ph}}$ is built from $S$, by gluing $S$ diagrams together
into new diagrams, through non time-ordered propagators. We classify the
most general way to gain unitarity by means of such operations, and point
out that a special solution \textquotedblleft minimizes\textquotedblright\
the time-ordering violation. We show that the scattering matrix $S_{\text{ph}%
}$ given by this solution coincides with the one obtained by turning the
would-be ghosts (and possibly some would-be physical particles) into purely
virtual particles (fakeons). We study tricks to descend and ascend in a
unique way among diagrams, and illustrate them in several examples: the
ascending chain\ from the bubble to the hexagon, at one loop; the box with
diagonal, at two loops; other diagrams, with more loops.

\vfill\eject

\section{Introduction}

\label{intro}\setcounter{equation}{0}

Unitarity is a fundamental principle of quantum field theory. It states that
the scattering matrix $S$ satisfies the unitarity equation $S^{\dag }S=1$.
Equivalently, the $T$ matrix defined by $S=1+iT$ satisfies the optical
theorem%
\begin{equation}
iT-iT^{\dag }=-T^{\dag }T.  \label{finalized}
\end{equation}%
The virtue of this formula is that it can be converted into Cutkosky-Veltman
identities \cite{cutkosky,veltman,thooft,veltman2}, which are diagrammatic
relations, satisfied by each diagram separately. The diagrams of $T$ are
built by means of the usual vertices and propagators. The diagrams of $%
T^{\dag }$ are built by means of the conjugate vertices and the conjugate
propagators. The diagrams of $T^{\dag }T$ have two sides, one for $T^{\dag }$
and one for $T$, separated by a \textquotedblleft cut\textquotedblright .
The product between $T^{\dag }$ and $T$ is rendered diagrammatically by
means of \textquotedblleft cut propagators\textquotedblright , which are on
shell and encode the physical contents of the theory.

Thus, while the matrix $T$ is given by the usual Feynman diagrams, the
identity (\ref{finalized}) involves the larger class of Cutkosky-Veltman
diagrams, which are also called \textquotedblleft cut
diagrams\textquotedblright . It was shown in ref. \cite{diagrammarMio} that
the identities obeyed by the \textquotedblleft skeletons\textquotedblright\
of the diagrams (where we ignore the integrals on the space components of
the loop momenta) split into independent \textit{spectral optical identities}%
, one for every (multi)threshold. The virtue of these relations is that they
are algebraic and relatively straightforward to manipulate. Moreover, they
provide the \textit{threshold decomposition} of a diagram, which can be used
to quantize the would-be ghosts, and possibly some would-be physical
particles, as \textit{purely virtual particles}, thereby projecting the
matrix $T$ onto a reduced matrix $T_{\text{ph}}$, which may be physically
acceptable even if $T$ is not.

Purely virtual particles, also called fake particles, or fakeons, are
defined by this new diagrammatics \cite{diagrammarMio}. The projection
allows us to remove degrees of freedom from the physical spectrum at all
energies, and satisfy the optical theorem in a manifest way. The main
application of the idea is the formulation of a consistent theory of quantum
gravity \cite{LWgrav}, which is observationally testable due to its
predictions in inflationary cosmology \cite{ABP}. At the phenomenological
level, fakeons evade common constraints that limit the employment of normal
particles (see \cite{PivaMelis} and references therein).

\medskip

In this paper, we study the scattering matrix of quantum field theory under
a new light. We inquire what transformations we can make on the usual $S$
matrix, which is defined by the time-ordered product, to turn it into a
different scattering matrix $S_{\text{ph}}$, possibly better suited to
describe the physics we observe in experiments. We end up by uncovering
purely virtual particles again, in an independent way. The results confirm
and upgrade the ones of \cite{diagrammarMio} and provide an alternative
understanding of the concept of purely virtuality.

\medskip

Unitarity is not an automatic consequence of the usual quantization
principle, and should not be taken for granted. For example, if we start
from a theory that contains fields with negative kinetic terms, and quantize
it as usual, we obtain a matrix $T$ that is not physically acceptable,
because it does not satisfy the unitarity equation (\ref{finalized}). Even
in that case, though, $T$ satisfies a mathematically useful identity, which
reads \cite{thooft,veltman2} 
\begin{equation}
iT-iT^{\dag }=-T^{\dag }CT  \label{interim}
\end{equation}%
and is called pseudounitarity equation, where $C$ is a diagonal matrix with
eigenvalues equal to 1, $-$1 and possibly 0 (if auxiliary fields are
present). Higher-derivative theories are typical examples of theories
satisfying (\ref{interim}) with $C\neq \mathbb{I}$.

Normally, the quantization ends with the $S$ matrix, which means that if $%
T=i-iS$ is physically unacceptable the theory is discarded. What if the
quantization did not end there? What if the derivation of $T$ were just the
first step of a longer, more elaborate quantization procedure? To make this
happen, we need a new quantization principle, equivalent to the old one
whenever the old one was successful, but possibly differing from it in every
other case. In particular, it should contemplate a second step, defined by
new diagrammatic rules, and a map $T\rightarrow T_{\text{ph}}$ from the
\textquotedblleft interim\textquotedblright\ matrix $T$ satisfying (\ref%
{interim}),\ to the \textquotedblleft finalized\textquotedblright ,
hopefully physical, matrix $T_{\text{ph}}$, satisfying (\ref{finalized}).

The first task is to classify all the possibilities we have to build a
unitary scattering matrix $S_{\text{ph}}=1+iT_{\text{ph}}$ from another
unitary scattering matrix, or from a pseudounitary one, $S=1+iT$. Since the
time-ordered product leads to the usual $S$ matrix, and leaves no room for
alternatives, the second part of the new quantization principle must be
based on non time-ordered products.

The set of solutions to the problem just stated is large, but a very special
one can be singled out among the others. It is the one that minimizes, so to
speak, the violation of the time ordering. A bonus is that it provides an
alternative way to uncover the physics of purely virtual particles.

\medskip

\textbf{The new quantization principle}

We briefly describe the new diagrammatics, and state the quantization
principle they lead to. If $\varphi $ denotes the fields, the diagrams of
the physical matrix $T_{\text{ph}}$ are built by means of the usual
vertices, the usual (time-ordered) free-field propagators 
\begin{equation}
\langle 0|\theta (x^{0}-y^{0})\varphi (x)\varphi (y)+\theta
(y^{0}-x^{0})\varphi (y)\varphi (x)|0\rangle _{0}  \label{Tprop}
\end{equation}%
and the non time-ordered free-field propagators 
\begin{equation}
\langle 0|\varphi (x)\varphi (y)|0\rangle _{0}.  \label{nonTprop}
\end{equation}%
The rules to build the diagrams with these ingredients are encoded into a
compact formula, which is eq. (\ref{ups}) of section \ref{key}, and an
iterative procedure to eliminate a certain arbitrariness $\Omega $ contained
in that same formula.

Curiously enough, the non time-ordered propagators (\ref{nonTprop})\
coincide with the cut propagators mentioned earlier. However, the cut
propagators are not used to build the $T$ diagrams, which define the
ordinary transition amplitudes. They appear in the Cutkosky-Veltman
diagrams, when we study the (pseudo)unitarity equations (\ref{finalized})
and (\ref{interim}) obeyed by $T$. Specifically, they connect $T^{\dag }$
and $T$ through the products appearing on the right-hand sides of those
equations. The first, crucial novelty of the new diagrammatics is that the
non time-ordered propagators (\ref{nonTprop}) are ingredients of the
diagrams that give the physical transition amplitudes, collected in $T_{%
\text{ph}}$. This way, the physical scattering matrix $S_{\text{ph}}=1+iT_{%
\text{ph}}$ is no longer dictated by the time-ordered product. The inclusion
of non time-ordered propagators multiplies the number of diagrams we have to
consider by a large factor. However, the new diagrams are not more difficult
than the usual ones, and their large number can be easily dealt with by
means of computer programs, like the popular ones used nowadays in
phenomenology \cite{calc}.

Briefly, the usual diagrams contributing to $T$ are glued together in
certain, prescribed ways, by means of the non time-ordered propagators, to
build the new diagrams, those of $T_{\text{ph}}$. A certain formula mapping
the standard matrix $T$ to the physical matrix $T_{\text{ph}}$, and a
certain procedure, guide the assembly of the new diagrams.

The new quantization principle is thus made of two parts. The first part
amounts to build the matrix $T$ as usual. The second part amounts to work
out the physical matrix $T_{\text{ph}}$ as explained. The map $T\rightarrow
T_{\text{ph}}$ is sometimes called \textquotedblleft
projection\textquotedblright , other times \textquotedblleft
reduction\textquotedblright , interchangeably.

\medskip

We show that the most general reduced matrix $T_{\Omega }$ built from $T$,
which obeys the unitarity equation, depends on an arbitrary anti-Hermitian
matrix $\Omega $. A special $\Omega $ is singled out by requiring that the
projection of a product diagram is equal to the product of the projected
factors, and the factorization survives basic diagrammatic operations. Due
to the violation of time ordering, this factorization requirement is
nontrivial. It amounts to assume that the violation is a \textquotedblleft
minimum\textquotedblright\ one, rather than the most brutal one: it is
confined inside non factorizable diagrams, which we call \textquotedblleft
prime\textquotedblright\ diagrams. The reason why we call it minimum
violation is that it does not seem possible to violate it less than this.
With this particular choice of $\Omega $, the large number of diagrams
collapses to an amount that is comparable to the one generated by the usual
time-ordered product.

We show that the reduced matrix $T_{\Omega }$ obtained this way coincides
with the matrix $T_{\text{ph}}$ of a theory of physical and purely virtual
particles, as given in ref. \cite{diagrammarMio}. Once we decide which
particles we want to quantize as physical and which ones we want to quantize
as purely virtual, $T_{\text{ph}}$ follows uniquely.

\medskip

We briefly mention the other results of the paper. We work out a number of
tricks to descend from bigger to smaller diagrams, but also ascend in a
unique way from smaller to bigger diagrams, and relate their $\Omega $
corrections and their threshold decompositions. We illustrate these
properties in various examples. At one loop, we study the ascending chain 
\begin{equation*}
\text{bubble}\rightarrow \text{triangle}\rightarrow \text{box}\rightarrow 
\text{pentagon}\rightarrow \text{hexagon.}
\end{equation*}%
At two loops, we focus on the first nontrivial arrangement, which is the box
diagram with diagonal. Classes of diagrams with arbitrarily many loops are
also discussed. Agreement with the formulas of \cite{diagrammarMio} is found
in every example.

A diagram may need an overall $\Omega $ correction, but also inherit $\Omega 
$ corrections from its subdiagrams. In all the cases we consider, nontrivial
corrections are present when the diagram is not prime, and also when it is
prime, but factorizes under the contraction of some internal legs. We find
that it is always possible to ascend through the threshold decompositions in
a unique way. We conjecture that these are general properties of the
physical matrix $T_{\text{ph}}$.

\medskip

Since the time-ordered product is not a physical principle, we should be
open to the possibility that the physical laws may break it, one way or
another. Purely virtual particles provide the most elegant and economic way
of implementing such a breaking. In physical applications, purely virtual
particles are expected to be massive, and generically heavy. For example,
one spin-2 purely virtual particle $\chi _{\mu \nu }$ of mass $m_{\chi }\sim 
$10$^{12-13}$GeV is enough to make sense of quantum gravity \cite{LWgrav}.
In that case, the violation of time ordering is restricted to distances $%
\lesssim 1/m_{\chi }$, and so is the violation of microcausality associated
with it (as well as the violation of microlocality, when $\chi _{\mu \nu }$
is integrated out). Tiny violations like these are not detectable in
realistic situations, even if we take into account the possibility of
boosting the systems. We also recall that purely virtual particles in curved
space lead to a sharp prediction for the tensor-to-scalar ratio $r$ in
primordial cosmology ($0.0004\lesssim r\lesssim 0.0035$ \cite{ABP}). The
first observational results on this are expected to become available in the
present decade \cite{CMBStage4}.

The results of this paper provide a quantization, in perturbation theory, of
any theory for which the usual Feynman diagram calculation of the $T$ matrix
gives a result satisfying (\ref{interim}). In particular, no assumption
about gauge invariance is used. This means that we can make sense of a very
wide class of theories usually thought unacceptable. Many workers in quantum
field theory believe that negative probability modes can only be removed if
the theory has a gauge invariance, and that the physical states are selected
through that symmetry. The construction of this paper provides
counterexamples to that belief.

Ultimately, the correctness of the new ideas must be proven by experiment,
for example by confirming the prediction for $r$, or the viability of
standard model extensions such as the one of \cite{Tallinn}. If theories
constructed with these methods turn out to be phenomenologically correct,
then we need to expand our orthodox ideas about fundamental physics. Further
insight on this aspect could come from the investigation of an important
open problem, that is to say, establish whether the $S$ matrix generated by
the prescription for purely virtual particles is the result of a Hamiltonian
evolution, as we normally understand it, or we need to relax the basic
axioms of quantum mechanics. Although this issue is beyond the scope of this
paper, it ought to be explored.

\medskip

The paper is organized as follows. In section \ref{key} we work out the most
general map that converts a matrix $T$ satisfying the (pseudo)unitarity
equation (\ref{interim}) into a reduced matrix $T_{\text{red}}$ that
satisfies the unitarity equation (\ref{finalized}). In section \ref%
{PVphysics} we outline the basic rules for the new diagrams. In sections \ref%
{treediagrams} and \ref{product}, we discuss the reduction $T\rightarrow T_{%
\text{red}}$ in the cases of tree and disconnected diagrams. In section \ref%
{oneloop} we study the simplest one-loop diagrams (bubble, triangle and
box). In section \ref{moreloop} we study diagrams with more loops, focusing
on the box with diagonal, which is the first truly new arrangement. In
section \ref{purevirtuality} we discuss a number of tools to have control on
pure virtuality, and relate smaller and bigger diagrams. In section \ref%
{ascension} we explain how to use those tricks to ascend and descend through
the diagrams and their threshold decompositions. In section \ref{rules} we
summarize the diagrammatic rules, and compare the main options (Feynman
diagrams, Cutkosky-Veltman diagrams, and the diagrams defined here), and
their uses. In section \ref{Symmetries} we show that the projection
preserves the global and local symmetries of a theory, the cancellation of
anomalies and the renormalizability. Section \ref{conclusions} contains the
conclusions. In appendix \ref{diagrammar20} we explain how to switch from
the scattering matrix to single diagrams, without loss of information, and
vice versa, to derive the diagrammatic identities. In appendix \ref{app2} we
prove some identities for product distributions, used in the paper.

\section{The key issue and its solution}

\label{key}\setcounter{equation}{0}

In this section we describe how to reduce the scattering matrix of a
possibly nonunitary theory to a unitary scattering matrix. We classify the
set of solutions without assuming physical inputs.

We decompose the usual $S$ matrix as%
\begin{equation}
S=1+V,  \label{sup}
\end{equation}%
where $V=iT$, and $T$ collects the common transition amplitudes, defined by
the time-ordered product. From now on, we refer to $V$, or, more generally,
the difference between a scattering matrix and the identity matrix, by
simply calling it \textquotedblleft amplitude\textquotedblright .

The unitarity of the $S$ matrix, i.e., the identity $S^{\dag }S=1$, gives
the identity%
\begin{equation}
V+V^{\dag }=-V^{\dag }V  \label{unit}
\end{equation}%
for the amplitude $V$. Various quantum field theories do not allow us to
prove an equation of this form right away. When the theory has fields with
negative kinetic terms, we can just prove a pseudounitarity equation%
\begin{equation}
V+V^{\dag }=-V^{\dag }CV,  \label{C}
\end{equation}%
where $C$ is some Hermitian matrix. We can diagonalize and normalize $C$ so
as to put it into the form 
\begin{equation}
\text{diag}(\overset{n_{+}}{\overbrace{1,\cdots ,1}},\overset{n_{-}}{%
\overbrace{-1,\cdots ,-1}},\overset{n_{0}}{\overbrace{0,\cdots ,0}}).
\label{CC}
\end{equation}%
The corresponding Fock space decomposition is written as $W=W_{+}\oplus
W_{-}\oplus W_{0}$, where $W$ is the total Fock space.

A quick derivation of (\ref{C}) from (\ref{unit}) goes on as follows. We
integrate out the auxiliary fields, for simplicity, and use $\hat{\varphi}$
to denote the fields that have negative kinetic terms. If, for a moment, we
change the signs of the $\hat{\varphi}$ propagators, we obtain a modified
theory that satisfies (\ref{unit}). Consider a diagram $G$ of the modified
theory, and the diagrammatic equation satisfied by it, generated by (\ref%
{unit}). If we multiply that equation by a factor $(-1)^{n_{G}}$, where $%
n_{G}$ denotes the number of the $\hat{\varphi}$ legs, we restore the
factors of the original theory in front of the propagators of the internal $%
\hat{\varphi}$ legs due to $V^{\dag }$ and $V$. However, the cut $\hat{%
\varphi}$ legs connecting $V^{\dag }$ and $V$ also get factors $(-1)$. This
converts the unit matrix $\mathbb{I}$ between $V^{\dag }$ and $V$ into the
matrix $C$, leading to formula (\ref{C}).

Our goal is to project the amplitude $V$ and the space $W$, so as to obtain
an equation like (\ref{unit}) from (\ref{C}), holding in a \textquotedblleft
physical\textquotedblright\ subspace $W_{\text{ph}}$ of $W$.

Let%
\begin{equation}
\Pi _{\text{ph}}=\text{diag}(\overset{N_{\text{ph}}}{\overbrace{1,\cdots ,1}}%
,\overset{N_{\text{pv}}}{\overbrace{0,\cdots ,0}}),  \label{A}
\end{equation}%
denote the projector onto $W_{\text{ph}}$, and $W=W_{\text{ph}}\oplus W_{%
\text{pv}}$ the corresponding $W$ decomposition, where $N_{\text{ph}%
}\leqslant n_{+}$, $N_{\text{pv}}=n_{+}+n_{-}+n_{0}-N_{\text{ph}}$. It is
enough to find a \textit{reduced} amplitude $V_{\text{red}}$ that solves the
equation%
\begin{equation}
V_{\text{red}}+V_{\text{red}}^{\dag }=-V_{\text{red}}^{\dag }\Pi _{\text{ph}%
}V_{\text{red}}.  \label{Thetaproj}
\end{equation}%
Indeed, this equation implies that the physical amplitude $V_{\text{ph}%
}\equiv \Pi _{\text{ph}}V_{\text{red}}\Pi _{\text{ph}}$ solves%
\begin{equation}
V_{\text{ph}}+V_{\text{ph}}^{\dag }=-V_{\text{ph}}^{\dag }V_{\text{ph}},
\label{unita}
\end{equation}%
so the physical $S$ matrix $S_{\text{ph}}\equiv \Pi _{\text{ph}}+V_{\text{ph}%
}$ satisfies $S_{\text{ph}}^{\dag }S_{\text{ph}}=\Pi _{\text{ph}}$. In other
words, if we manage to solve (\ref{Thetaproj}), we achieve unitarity in the
subspace $W_{\text{ph}}$. Then we can legitimately claim that $V_{\text{ph}}$
is the physical amplitude, and $W_{\text{ph}}$ is physical space\ of the
theory.

Summarizing, our goal is to find the most general solution $V_{\text{red}}$
of (\ref{Thetaproj}), given the physical projector $\Pi _{\text{ph}}$.

We can generalize the problem a little bit with no effort: given Hermitian
matrices $C$ and $A$, and given a matrix $V$ that satisfies (\ref{C}), we
want to find the most general solution $V_{\text{red}}$ of the equation%
\begin{equation}
V_{\text{red}}+V_{\text{red}}^{\dag }=-V_{\text{red}}^{\dag }AV_{\text{red}}.
\label{Theta}
\end{equation}%
This is a merely mathematical problem about matrices, and $A$, $C$ do not
need to be projectors or linear combinations of projectors. For convenience,
we write\ 
\begin{equation}
B=C-A.  \label{CAB}
\end{equation}

A particular solution of (\ref{Theta}) is $V_{\text{red}}=V_{0}$, where%
\begin{equation}
V_{0}=\left( 1+\frac{1}{2}VB\right) ^{-1}V.  \label{Thetasol}
\end{equation}%
This formula can be understood recursively as%
\begin{equation*}
V_{0}=V-\frac{1}{2}VBV_{0}=V-\frac{1}{2}VBV+\frac{1}{4}VBVBV+\cdots .
\end{equation*}%
Note that, despite its appearance, the solution is left-right symmetric,
since we can also write 
\begin{equation*}
V_{0}=V-\frac{1}{2}V_{0}BV=V-\frac{1}{4}V_{0}BV-\frac{1}{4}VBV_{0}.
\end{equation*}

The proof that (\ref{Thetasol}) solves $V_{0}+V_{0}^{\dag }=-V_{0}^{\dag
}AV_{0}$ follows straightforwardly from the identities 
\begin{equation}
V^{\dag }=-V\left( 1+CV\right) ^{-1},\qquad V_{0}^{\dag }=-V\left( 1+AV+%
\frac{1}{2}BV\right) ^{-1},  \label{iden}
\end{equation}
which are implied by (\ref{C}) and then (\ref{Thetasol}).

Starting from the particular solution (\ref{Thetasol}), we can write the
most general solution $V_{\text{red}}$ as%
\begin{equation}
V_{\text{red}}=V_{0}+\Omega _{0}.  \label{upsi}
\end{equation}%
In order to fulfill (\ref{Thetasol}), the matrix $\Omega _{0}$ must satisfy
the equation%
\begin{equation}
\Omega _{0}+\Omega _{0}^{\dag }=-V_{0}^{\dag }A\Omega _{0}-\Omega _{0}^{\dag
}AV_{0}-\Omega _{0}^{\dag }A\Omega _{0}.  \label{Omega}
\end{equation}%
As before, we can solve this equation recursively in powers of $V_{0}$,
starting from an arbitrary anti-Hermitian matrix $\tilde{\Omega}$ that is at
least of order $V_{0}$. Indeed, if we write $\Omega _{0}=\tilde{\Omega}%
+\Delta \Omega $, with $\tilde{\Omega}+\tilde{\Omega}^{\dag }=0$ and assume
that $\Delta \Omega $ is of higher order in the expansion, we obtain the
equation%
\begin{equation*}
\Delta \Omega +\Delta \Omega ^{\dag }=-\Omega _{0}^{\dag }A\Omega
_{0}-V_{0}^{\dag }A\Omega _{0}-\Omega _{0}^{\dag }AV_{0},
\end{equation*}%
which can be solved iteratively as claimed. This also proves that the matrix 
$\tilde{\Omega}$ parametrizes the most general solution of (\ref{Omega}).

It is simple to show that the explicit solution $\Omega _{0}$ of (\ref{Omega}%
), and its inverse, are%
\begin{equation}
\Omega _{0}=\left( 1-\frac{1}{2}\tilde{\Omega}A\right) ^{-1}\tilde{\Omega}%
\left( 1+AV_{0}\right) ,\qquad \tilde{\Omega}=\Omega _{0}\left( 1+\frac{1}{2}%
A\Omega _{0}+AV_{0}\right) ^{-1}.  \label{forma}
\end{equation}%
The proof follows by writing (\ref{Omega}) in the form $E+E^{\dag }=0$, where%
\begin{equation*}
E=\Omega _{0}^{\dag }\left( 1+\frac{1}{2}A\Omega _{0}+AV_{0}\right) .
\end{equation*}%
Using the expression (\ref{forma}) of $\Omega _{0}$, and $\tilde{\Omega}%
^{\dag }=-\tilde{\Omega}$, we see that the matrix%
\begin{equation*}
E=-\left( 1+V_{0}^{\dag }A\right) \left( 1+\frac{1}{2}\tilde{\Omega}A\right)
^{-1}\tilde{\Omega}\left( 1-\frac{1}{2}A\tilde{\Omega}\right) ^{-1}\left(
1+AV_{0}\right)
\end{equation*}%
is indeed anti-Hermitian.

The solution (\ref{upsi}), with $V_{0}$ given in (\ref{Thetasol}) and $%
\Omega _{0}$ given in (\ref{forma}), needs some rearrangement, since it is
not written in a manifestly left-right symmetric form. The symmetrization
can be obtained by redefining $\tilde{\Omega}$ within the realm of its own
arbitrariness. Define%
\begin{equation}
\Omega =\left( 1+V_{0}A\right) ^{-1/2}\tilde{\Omega}\left( 1+AV_{0}\right)
^{1/2}.  \label{Omeg}
\end{equation}%
It is easy to prove that $\Omega $ is anti-Hermitian. To this purpose, note
that, since $V_{\text{red}}=V_{0}$ solves (\ref{Theta}), we have the formula%
\begin{equation*}
V_{0}^{\dag }=-V_{0}\left( 1+AV_{0}\right) ^{-1}.
\end{equation*}%
Equipped with (\ref{Omeg}), the first expression of (\ref{forma}) can be
recast into the form%
\begin{equation*}
\Omega _{0}=\left( 1+V_{0}A\right) ^{1/2}\left( 1-\frac{1}{2}\Omega A\right)
^{-1}\Omega \left( 1+AV_{0}\right) ^{1/2},
\end{equation*}%
which is manifestly left-right symmetric. Then, so is (\ref{upsi}).

\medskip

Going back to formula (\ref{upsi}) and summarizing the results we have found
so far, we have proved that, given a matrix $V$ that satisfies%
\begin{equation*}
V+V^{\dag }=-V^{\dag }AV-V^{\dag }BV,
\end{equation*}%
where $A$ and $B$ are arbitrary Hermitian matrices, the most general matrix $%
V_{\text{red}}$ that satisfies 
\begin{equation*}
V_{\text{red}}+V_{\text{red}}^{\dag }=-V_{\text{red}}^{\dag }AV_{\text{red}}
\end{equation*}%
and coincides with $V$ up to corrections of higher orders in $V$ itself, is $%
V_{\text{red}}=V_{\Omega }(A,B)$, where%
\begin{equation}
\boxed{\begin{array}{l} \quad V_\Omega(A,B) =\left( 1+\frac{1}{2}V B\right)
^{-1}V \\ \quad\qquad\qquad+\left( 1+\left( 1+\frac{1}{2}V B\right) ^{-1}V
A\right) ^{1/2}\left( 1-\frac{1}{2}\Omega A\right) ^{-1}\Omega \left( 1+AV
\left( 1+\frac{1}{2}BV \right) ^{-1}\right) ^{1/2}\!\!\!\!\!\quad
\label{ups} \end{array}}
\end{equation}%
and $\Omega $ is an arbitrary anti-Hermitian matrix, to be considered of
order two in $V$, or higher. Formula (\ref{ups}) is the key formula of the
paper.

\medskip

An interesting case is when the physical space $W_{\text{ph}}$ is just made
of the vacuum state $|0\rangle $. Then $A$ is $|0\rangle \langle 0|$ and $B$
is $C-|0\rangle \langle 0|$. If $C$ has the form (\ref{CC}), the solution $%
V_{\Omega }(|0\rangle \langle 0|,C-|0\rangle \langle 0|)$ can be used to
remove the whole on-shell contents of the diagrams, and describe the
situation where every particle is rendered purely virtual. To achieve this
goal, $\Omega $ must be determined so as to remove any residual on-shell
contributions. As we are going to show in the next sections, this is a
nontrivial task, but has a well defined answer. Unfortunately, the answer is
not just $\Omega =0$. Indeed, the $\Omega =0$ solution 
\begin{equation}
\mathring{V}\equiv V_{0}(|0\rangle \langle 0|,C-|0\rangle \langle 0|)\equiv
\left( 1+\frac{1}{2}V(C-|0\rangle \langle 0|)\right) ^{-1}V  \label{primeT}
\end{equation}%
turns out to be correct only in a certain subset of simpler diagrams. In
general, a nonvanishing $\Omega $ is to be expected. We will show that the
solution $V_{\Omega }(|0\rangle \langle 0|,C-|0\rangle \langle 0|)$,
equipped with the right $\Omega $, provides an alternative way to make the
threshold decomposition of \cite{diagrammarMio}.

By affinity with the notion of prime number, we say that a diagram is 
\textit{prime} if it cannot be factorized as a nontrivial product of smaller
diagrams, in momentum space. We show that, for arbitrary $A$, $\Omega $ can
be chosen to make $V_{\Omega }$ obey the following factorization property:
the projection of a non prime diagram is the product of the projections of
its prime factors, and the factorization survives basic operations of ascent
and descent among diagrams. This is a nontrivial requirement, in the realm
of non time-ordered products. We also show that the amplitude $V_{\Omega }$
determined by this $\Omega $ gives precisely the diagrams of physical and
purely virtual particles, as per ref. \cite{diagrammarMio}. We identify such
a\ $V_{\Omega }$ with the physical amplitude $V_{\text{ph}}$.

We first proceed by explicit examples, then gather the lessons we learn
along the way.

\section{Diagrams: the old and the new}

\label{PVphysics}\setcounter{equation}{0}

In this section we lay out the rules to build the new diagrams, and compare
them with the Feynman rules.

So far, we have been merely playing with matrices: the theorems proved in
the previous section hold under assumptions that are more general than the
ones we need for physical applications. To move forward towards the physics,
let $\Phi =\{\varphi ,\chi \}$ collect all the fields, which include the
physical fields $\varphi $ and the fields $\chi $ we want to project away
(for one reason or another). The physical subspace $W_{\text{ph}}$ contains
the vacuum state $|0\rangle $ and the states that are built by means of the $%
\varphi $ creation operators, but no $\chi $ creation operators. The
complementary subspace $W_{\text{pv}}$ contains the states that are built by
means of at least one $\chi $ creation operator. The idea is that a single
excitation due to the fields that we want to get rid of is sufficient to
drop the whole state from the physical spectrum.

From our definition (\ref{sup}), it follows that the amplitude $V$ collects
the usual Feynman diagrams. In operatorial notation,%
\begin{equation}
V=\mathcal{T}\exp \left( -i\int_{-\infty }^{+\infty }H_{I}(t)\mathrm{d}%
t\right) -1,  \label{upsop}
\end{equation}%
where $H_{I}$ is the interaction Hamiltonian and $\mathcal{T}$ denotes the
time-ordered product.

The $V$ diagrams are defined by the usual Feynman rules. In particular, the $%
\varphi $ free-field propagators are the time-ordered ones, given by the
Feynman $i\epsilon $ prescription. For scalars $\varphi $, we have%
\begin{equation}
\langle 0|\mathcal{T}\varphi (x)\varphi (y)|0\rangle _{0}=\int \frac{\mathrm{%
d}^{4}p}{(2\pi )^{4}}\mathrm{e}^{-ip(x-y)}\frac{i}{p^{2}-m^{2}+i\epsilon }.
\label{spropa}
\end{equation}%
The propagators of the $\chi $ fields are the same, apart from the
possibility of being multiplied by minus signs. Thus, $\chi $ scalars may
have propagators 
\begin{equation}
\langle 0|\mathcal{T}\chi (x)\chi (y)|0\rangle _{0}=\int \frac{\mathrm{d}%
^{4}p}{(2\pi )^{4}}\mathrm{e}^{-ip(x-y)}\frac{(\pm i)}{p^{2}-m^{2}+i\epsilon 
},  \label{prop}
\end{equation}%
The 0 eigenvalues of the matrix $C$ of (\ref{CC}) correspond to auxiliary
fields $\chi $. From now on, we assume that they are integrated away.

The right-hand side $-V^{\dag }CV$ of formula (\ref{C}) collects the usual
Cutkosky-Veltman diagrams, which are graphically rendered by means cut
diagrams. The cut is unique, and represents the matrix $C$ separating $%
V^{\dag }$ from $V$. While the $V$ diagrams are time-ordered, and the $%
V^{\dag }$ diagrams are anti-time-ordered, the product $V^{\dag }CV$ is just
a plain (non time-ordered) product of field operators. This means that, when
we apply Wick's theorem, the Wick contraction between a field $\Phi (x)$
that belongs to a conjugate vertex $\bar{v}_{1}$ of $V^{\dag }$, and a field 
$\Phi (y)$ that belongs to an ordinary vertex $v_{2}$ of $V$ is the non
time-ordered two-point function%
\begin{equation}
\langle 0|\Phi (x)\Phi (y)|0\rangle _{0}=\pm \int \frac{\mathrm{d}^{3}%
\mathbf{p}}{(2\pi )^{3}2\omega }\mathrm{e}^{-ip(x-y)}=\pm \int \frac{\mathrm{%
d}^{4}p}{(2\pi )^{4}}(2\pi )\theta (p^{0})\delta (p^{2}-m^{2})\mathrm{e}%
^{-ip(x-y)},  \label{cutpropa}
\end{equation}%
where the sign is $+$ or $-$ according to the sign of the $C$ eigenvalue
associated with $\Phi $, as in (\ref{spropa}) and (\ref{prop}). In momentum
space these \textquotedblleft cut propagators\textquotedblright\ are thus%
\begin{equation}
\pm (2\pi )\theta (p^{0})\delta (p^{2}-m^{2}).  \label{cutprop}
\end{equation}

\medskip

Now we describe the diagrams of the reduced amplitude $V_{\Omega }$. We
assume that $A$ is the projector $\Pi _{\text{ph}}$ onto the physical space $%
W_{\text{ph}}$, $C$ has the form (\ref{CC}), and $B=C-A$. The $V_{\Omega }$
diagrams follow from formula (\ref{ups}), by expanding the right-hand side
in powers of $V$. Each term of the expansion is graphically represented as a
cut diagram, multiplied by a coefficient inherited from the expansion
itself. We must distinguish two types of cuts, standing for the matrices $A$
and $B$. A $V_{\Omega }$ diagram may contain an arbitrary number of such
cuts.

It is convenient to draw the cuts as vertical lines and place the vertices
in the strips between pairs of consecutive cuts, and in the half planes
located at the sides. Every strip, or half plane, must contain at least one
vertex.

The vertices and the uncut propagators of the $V_{\Omega }$ diagrams
coincide with those of $V$. In particular, no conjugate vertices, nor
conjugate propagators, are involved. As above, a line crossed by a cut
stands for the propagator (\ref{cutprop}), the energy flowing conventionally
from the right to the left. The cut propagator contributes to $A$, or $B$,
depending on whether the field $\Phi $ of (\ref{cutpropa}) belongs to the
subset of physical fields $\varphi $, or the subset of fields $\chi $ we
want to project away. A cut $A$ can only cut $\varphi $ legs, while a cut $B$
must cut at least one $\chi $ leg.

The diagrams we consider in the examples of the next sections have a
different particle on each internal leg. In appendix \ref{diagrammar20} we
show that we can always enlarge the set of fields enough to fit this
arrangement, with no gain nor loss of information, by means of a
Pauli-Villars trick \cite{PV}. Diagrams with internal legs associated with
the same $\varphi $, or the same $\chi $, can be seen as particular cases.

We distinguish the various fields $\varphi $ and $\chi $ by means of indices 
$i$, and write $\Phi =\{\varphi _{i},\chi _{i}\}$. It is easy to show that
the diagrams where every $\chi _{i}$ appears an even number of times to the
right (left) of a $B$ cut vanish. Indeed, the Wick contraction makes the
creation and annihilation operators of all the $\chi $ fields disappear to
the right (left) of that cut. This means that only creation and annihilation
operators of physical fields $\varphi _{i}$ act on $|0\rangle $ ($\langle 0|$%
) before $B$. Since $B$ vanishes on the physical space $W_{\text{ph}}$, we
obtain something like $\langle 0|($physical fields $\varphi _{i})B$, or $B($%
physical fields $\varphi _{i})|0\rangle $, which vanish as well.

Once we decide what theory we want to build, we have the physical space $W_{%
\text{ph}}$, and know what fields $\chi $ we want to project away. This
means that we have the matrix $A$, which is the projector onto $W_{\text{ph}}
$, as well as the matrix $B$, which is equal to $C-A$. At that point, we are
ready to study the $V_{\Omega }$ diagrams encoded in (\ref{ups}). Although
they are a large number, there is no difficulty to list them by means of
computer software.

What can we obtain with a generic $\Omega $? In principle, anything we want.
We can even jump from the $S$ matrix of one theory, say the $\varphi ^{4}$
theory, to the $S$ matrix of a completely different theory, say the standard
model. The identities proved in the previous section are general properties
of matrices, with no constraints from physics. In physical applications, $%
\Omega $ cannot be completely arbitrary. For example, it should be at least $%
\mathcal{O}(V^{2})$, as mentioned right after formula (\ref{ups}). Moreover,
it should not change the basic contents of the theory. This requirement can
be phrased more precisely by stating that: it should not change the
Euclidean version of the theory; equivalently, it should not change the
zeroth level of the threshold decomposition of diagrams (see the beginning
of section \ref{oneloop}). Finally, $\Omega $ cannot introduce singularities
that are not present in the Feynman diagrams (such as new thresholds, or new
types of singular behaviors around existing thresholds).

A particular solution $V_{\Omega }$ must give the diagrammatics of physical
and purely virtual particles, derived in ref. \cite{diagrammarMio}. Indeed,
the diagrams defined there also solve the problem of building a unitary
matrix $T_{\text{ph}}$ out of the ordinary (pseudo)unitary matrix $T$. Thus,
there must exist an $\Omega $ that makes the $V_{\Omega }$ diagrams coincide
with those of \cite{diagrammarMio}. Unfortunately, such an $\Omega $ is not
just $\Omega =0$, nor something comparably simple, but must be worked out
iteratively. The examples studied in the next sections tell us how, and make
us appreciate what makes the solution of \cite{diagrammarMio} so special.

The operations we have described are not all straightforward, so we must
spend some time to describe them in detail, starting from the connected tree
diagrams and the product diagrams, to conclude with the loop diagrams. We
mostly work with plus signs in front of the propagators (\ref{prop}) and (%
\ref{cutprop}), which means $C=\mathbb{I=}\sum_{n}|n\rangle \langle n|$,
where $|n\rangle $ is an orthonormal basis of states. The other cases are
easily obtained by flipping overall signs in front of formulas and
identities.

\section{Tree diagrams}

\label{treediagrams}\setcounter{equation}{0}

In this section and the next one, we apply the results of section \ref{key}
in relatively simple cases, which, however, show some surprises. This helps
us illustrate the meaning of the various formulas and their ingredients.

The simplest example is the\ free propagator. We take the interaction\
Lagrangian 
\begin{equation*}
\mathcal{L}_{I}=K_{1}\varphi +K_{2}\varphi =-H_{I},
\end{equation*}%
where $\varphi $ denotes a scalar field of mass $m$ and standard propagator (%
\ref{spropa}), while $K_{1}$ and $K_{2}$ are external sources.
Differentiating (\ref{upsop}) with respect to $iK$ once for each source and
setting the sources to zero afterward, we get%
\begin{equation*}
\left. \frac{\delta ^{2}V}{i\delta K_{1}(x)i\delta K_{2}(y)}\right\vert
_{K=0}=\mathcal{T}\varphi (x)\varphi (y).
\end{equation*}%
Averaging on the vacuum state, we obtain the propagator (\ref{spropa}).

The theory is unitary (it is just a free field theory), so $C=\mathbb{I}%
\equiv \sum_{n}|n\rangle \langle n|$. Moreover, $\sum_{n}\langle 0|\varphi
(x)|n\rangle \langle n|\varphi (y)|0\rangle =\langle 0|\varphi (x)\varphi
(y)|0\rangle $. In a product such as $VCV$, the Wick contraction between a $%
\varphi $ due to the left $V$ and a $\varphi $ due to the right $V$ is just
a product of field operators, with no time ordering, which gives the cut
propagator (\ref{cutprop}), with energy conventionally flowing from the
right to the left.

Now, choose $A=|0\rangle \langle 0|$, so $B=\mathbb{I-}|0\rangle \langle 0|$%
. Let us assume, for the moment, that the $\Omega $ correction vanishes.
Then, the amplitude $V_{\Omega }$ is given by (\ref{Thetasol}):%
\begin{equation}
V_{0}=\left( 1+\frac{1}{2}VB\right) ^{-1}V=V-\frac{1}{2}VBV+\frac{1}{4}VBVBV+%
\mathcal{O}(V^{4}).  \label{approx}
\end{equation}%
We can understand the meaning of this expression by concentrating on the
first correction, $-V\!BV/2$. Differentiating (\ref{approx}) with respect to 
$iK_{1}$ and $iK_{2}$, and setting the sources to zero afterward, we obtain%
\begin{eqnarray*}
\left. \frac{\delta ^{2}V_{0}}{i\delta K_{1}(x)i\delta K_{2}(y)}\right\vert
_{K=0} &=&\left. \frac{\delta ^{2}V}{i\delta K_{1}(x)i\delta K_{2}(y)}%
\right\vert _{K=0} \\
&&-\frac{1}{2}\left[ \frac{\delta V}{i\delta K_{1}(x)}B\frac{\delta V}{%
i\delta K_{2}(y)}+\frac{\delta V}{i\delta K_{2}(y)}B\frac{\delta V}{i\delta
K_{1}(x)}\right] _{K=0}.
\end{eqnarray*}%
Averaging on the vacuum state, we find%
\begin{equation*}
\langle 0|\mathcal{T}\varphi (x)\varphi (y)|0\rangle _{0}-\frac{1}{2}\langle
0|\varphi (x)\varphi (y)|0\rangle _{0}-\frac{1}{2}\langle 0|\varphi
(y)\varphi (x)|0\rangle _{0},
\end{equation*}%
that is to say, after Fourier transform,%
\begin{equation}
\frac{i}{p^{2}-m^{2}+i\epsilon }-\frac{1}{2}(2\pi )\theta (p^{0})\delta
(p^{2}-m^{2})-\frac{1}{2}(2\pi )\theta (-p^{0})\delta (p^{2}-m^{2})=\mathcal{%
P}\frac{i}{p^{2}-m^{2}},  \label{PVprop}
\end{equation}%
where $\mathcal{P}$ denotes the Cauchy principal value. The projected free
propagator is just the principal value, which contains no on-shell part.
This is precisely the free propagator of a purely virtual particle \cite%
{diagrammarMio}. Thus, the correction to $V$ subtracts the on-shell part of
the Feynman propagator, and renders the particle described by $\varphi $
purely virtual.

In passing, we recall that the propagator (\ref{PVprop}) cannot be used as
such inside loop diagrams: not surprisingly, a non time-ordered product must
be worked out on a diagram by diagram basis. Thus, the result (\ref{PVprop})
is not sufficient to claim that we are dealing with purely virtual articles:
it is just the first hint.

\subsection{Two propagators}

Now we consider a tree diagram made of two adjacent propagators, which we
denote by means of the symbol $\wedge $. We take the interaction Lagrangian 
\begin{equation*}
\mathcal{L}_{I}=K_{1}\varphi _{1}+K_{12}\varphi _{1}\varphi
_{2}+K_{2}\varphi _{2}=-H_{I},
\end{equation*}%
where $\varphi _{j}$ denote scalar fields with diagonalized kinetic terms,
masses $m_{j}$ and standard propagators%
\begin{equation*}
i\mathcal{F}_{j}\equiv \frac{i}{p^{2}-m_{j}^{2}+i\epsilon },
\end{equation*}%
$K_{i}$ and $K_{12}$ being external sources. For the moment, we consider the
case where we project both $\varphi _{1}$ and $\varphi _{2}$ away. Then we
have, again, $C=\mathbb{I}=\sum_{n}|n\rangle \langle n|$, $A=|0\rangle
\langle 0|$, $B=\mathbb{I-}|0\rangle \langle 0|$.

If we differentiate $V_{\Omega }$, (\ref{approx}) and identities like (\ref%
{C}) and (\ref{Theta}) once with respect to $iK_{1}(x_{1})$, $iK_{2}(x_{2})$
and $iK_{12}(x)$, and set the sources to zero afterward, we can study the
correlation function $\langle 0|\mathcal{T}\varphi _{1}(x_{1})\varphi
_{1}(x)\varphi _{2}(x)\varphi _{2}(x_{2})|0\rangle $ and its projection. We
start again from $\Omega =0$. 
\begin{figure}[t]
\begin{center}
\includegraphics[width=16truecm]{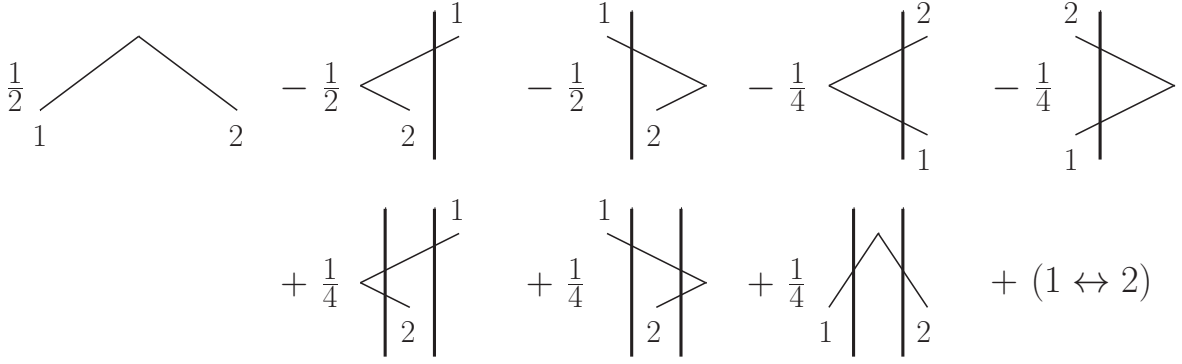}
\end{center}
\caption{{}Tree diagram with two propagators}
\label{DoubleProp}
\end{figure}

Denote the matrix $B$ by means of a vertical bar, standing for a cut across
which the energy conventionally flows from the right to the left. We can
drop $\mathcal{O}(V^{4})$ in formula (\ref{approx}), because it does not
contribute here. We remain with diagrams that have two, one and zero cuts.

We have to distribute the vertices of $\mathcal{L}_{I}$ in between the
vertical bars, as well as to the left and to the right of them, in all
possible ways. We must also include the exchanges of $\varphi _{1}$ and $%
\varphi _{2}$, and pay attention to the fact that each zone should contain
at least one vertex. The diagrams we obtain are shown in fig. \ref%
{DoubleProp}. It is understood that the vertices are equal to unity.

What is the meaning of a leg that is cut twice? Nothing particular, just the
propagation of a free particle in the stripe between two cuts. We have seen
above that a cut is a missing time ordering: when a field to the left of a
cut is contracted with a field to the right of the cut, we have the non
time-ordered propagator (\ref{cutprop}). It follows that two cuts on the
same line are the same as one cut.

Collecting the various contributions, we obtain 
\begin{equation}
V_{0}(\wedge ,\text{PV}^{2})=(i\mathcal{P}_{1})(i\mathcal{P}_{2})-\Delta
_{1}^{+}\Delta _{2}^{-}-\Delta _{1}^{-}\Delta _{2}^{+},  \label{resdo}
\end{equation}%
where the subscripts $1$ and $2$ refer to the legs $\varphi _{1}$ and $%
\varphi _{2}$. We have defined%
\begin{equation*}
\mathcal{P}_{i}=\mathcal{P}\frac{1}{p_{i}^{2}-m_{i}^{2}},\qquad \Delta
_{i}^{\pm }=\pi \theta (\pm p_{i}^{0})\delta (p_{i}^{2}-m_{i}^{2}),
\end{equation*}%
where $p_{i}^{\mu }=(p_{i}^{0},\mathbf{p}_{i})$ is the momentum of the $i$th
leg, flowing from the right to the left with respect to the ordering $x_{1}$-%
$x$-$x_{2}$, and $\omega _{i}=\sqrt{\mathbf{p}_{i}^{2}+m_{i}^{2}}$ is the $i$%
th frequency. Here and below, we use the notations $V_{0}$ and $V_{\Omega }$
with a different meaning with respect to before. Specifically, they stand
for the derivatives of the matrices $V_{0}$ and $V_{\Omega }$ of (\ref%
{Thetasol}) and (\ref{ups}) with respect to the sources $K$. Their arguments
are the type of diagram we are considering (here $\wedge $) and the types of
particles propagating inside.

We see that the result (\ref{resdo}) is not the product $(i\mathcal{P}_{1})(i%
\mathcal{P}_{2})$ of the projected propagators of the two legs. This means
two things:\ that the projection of a product diagram is not the product of
the projected factors; that we do not get the result predicted by the
diagrammatics of purely virtual particles, as per \cite{diagrammarMio}.

Both these issues can be solved by advocating a nonvanishing matrix $\Omega $%
, as allowed by formula (\ref{ups}). For our purposes, that formula can be
truncated to%
\begin{equation*}
V_{\Omega }\rightarrow V-\frac{1}{2}VCV+\frac{1}{4}VCVCV+\Omega .
\end{equation*}%
Besides dropping $\mathcal{O}(V^{4})$, which cannot contribute here, we have
also dropped the terms containing $A=|0\rangle \langle 0|$. Indeed, any
diagram with an $A$ cut is disconnected, because it contains some $%
V|0\rangle \langle 0|V$, while the diagram we are considering is connected
(and so are its cut versions).

If we set%
\begin{equation}
\left. \frac{\delta ^{3}\Omega }{i\delta K_{1}i\delta K_{2}i\delta K_{12}}%
\right\vert _{K=0}=\Delta _{1}^{+}\Delta _{2}^{-}+\Delta _{1}^{-}\Delta
_{2}^{+},  \label{delta12}
\end{equation}%
we cancel the last two terms of (\ref{resdo}) and obtain the desired result%
\begin{equation}
V_{\Omega }(\wedge ,\text{PV}^{2})=(i\mathcal{P}_{1})(i\mathcal{P}_{2}).
\label{ffres}
\end{equation}%
The correction (\ref{delta12}) is indeed generated by an anti-Hermitian
contribution to the matrix $\Omega $.

Formula (\ref{ffres}) is what we wanted:\ the result factorizes and
coincides with the one predicted by having purely virtual particles on the
internal legs. We thus learn that a possible role of $\Omega $ is to convert
the result to a better diagrammatic form, since the matrix formula (\ref{ups}%
) is not constrained to have a satisfactory one.

Assume now that one particle, say particle 1, is physical and the second
particle needs to be quantized as purely virtual. Recall that the matrix $A$
projects onto the physical subspace, made by the states built with $\varphi
_{1}$, while $B$ projects onto the complementary subspace. We must reinstate
all the contributions of the diagrams of fig. \ref{DoubleProp}, where the
left side, or the right side, of any $B$ cut are physical. This happens: $i$%
) when they contain no fields $\varphi _{2}$, and $ii$) when they contain
two fields $\varphi _{2}$ (which are going to disappear after Wick
contraction). Specifically, we must restore the 2nd, 3rd, 6th, 7th and 8th
diagram, plus the one obtained from the 8th by exchanging the legs 1 and 2.
The sum of these diagrams is%
\begin{equation*}
-(i\mathcal{F}_{2})(\Delta _{1}^{+}+\Delta _{1}^{-})+(\Delta _{1}^{+}+\Delta
_{1}^{-})(\Delta _{2}^{+}+\Delta _{2}^{-})=-(i\mathcal{P}_{2})(\Delta
_{1}^{+}+\Delta _{1}^{-}).
\end{equation*}%
Subtracting from (\ref{resdo}), we obtain%
\begin{equation}
V_{0}(\wedge ,\text{Ph-PV})=(i\mathcal{F}_{1})(i\mathcal{P}_{2})-\Delta
_{1}^{+}\Delta _{2}^{-}-\Delta _{1}^{-}\Delta _{2}^{+},  \label{resdoph}
\end{equation}%
at $\Omega =0$. Again, the result is not the one dictated by a theory of
physical and purely virtual particles. However, it becomes the desired one,
as soon as we choose the same $\Omega $ as in (\ref{delta12}), which
subtracts the last two terms. We finally obtain the factorized result%
\begin{equation}
V_{\Omega }(\wedge ,\text{Ph-PV})=(i\mathcal{F}_{1})(i\mathcal{P}_{2}).
\label{PhPV}
\end{equation}

\subsection{Three propagators}

\begin{figure}[t]
\begin{center}
\includegraphics[width=16truecm]{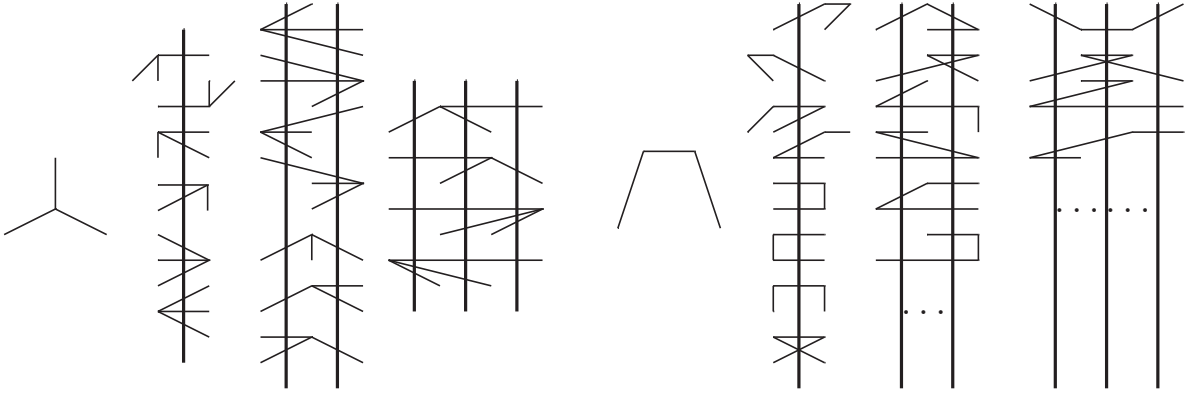}
\end{center}
\caption{Examples of tree diagrams with three propagators}
\label{TripleProp}
\end{figure}
Now we study two tree diagrams with three propagators. We have contributions
from cut diagrams that contain up to three cuts, shown in fig. \ref%
{TripleProp}.

In the first example, which we denote by \scalebox{1.5}{$\perp$}, the three
lines meet at the same point, so we take%
\begin{equation*}
\mathcal{L}_{I}=K_{1}\varphi _{1}+K_{2}\varphi _{2}+K_{3}\varphi
_{3}+K_{123}\varphi _{1}\varphi _{2}\varphi _{3},
\end{equation*}%
differentiate once with respect to each source (times $i$) and then set the
sources to zero. It is easy to show that, if $A=|0\rangle \langle 0|$, $B=%
\mathbb{I-}|0\rangle \langle 0|$, and $\Omega $ is chosen to be zero,
formula (\ref{Thetasol}) for $V_{0}$ gives 
\begin{equation}
V_{0}\big(\scalebox{1.5}{$\perp$},\text{PV}^{3}\big)=(i\mathcal{P}_{1})(i%
\mathcal{P}_{2})(i\mathcal{P}_{3})-\left[ (i\mathcal{P}_{1})(\Delta
_{2}^{+}\Delta _{3}^{+}+\Delta _{2}^{-}\Delta _{3}^{-})+\text{ cyclic
permutations}\right] ,  \label{resdokid}
\end{equation}%
where the energy flows are oriented towards the common vertex. The terms in
between the square brackets are then subtracted by means of $\Omega $. The
reduced amplitude $V_{\Omega }$ finally gives the non-amputated vertex of
three purely virtual particles,%
\begin{equation}
V_{\Omega }\big(\scalebox{1.5}{$\perp$},\text{PV}^{3}\big)=(i\mathcal{P}%
_{1})(i\mathcal{P}_{2})(i\mathcal{P}_{3}),  \label{fffres}
\end{equation}%
as desired.

An interesting configuration is the one where one leg, say $\varphi _{3}$,
is physical, while the other two need to be quantized as purely virtual. At $%
\Omega =0$ we must discard the diagrams of (\ref{ups}) where a $B$ cut
crosses only the physical leg $\varphi _{3}$. We obtain%
\begin{eqnarray}
&&(i\mathcal{P}_{1})(i\mathcal{P}_{2})(i\mathcal{F}_{3})-\left[ (i\mathcal{P}%
_{1})(\Delta _{2}^{+}\Delta _{3}^{+}+\Delta _{2}^{-}\Delta _{3}^{-})+\text{
cyclic permutations}\right]  \notag \\
&&\qquad \qquad \qquad -(\Delta _{1}^{+}\Delta _{2}^{+}+\Delta
_{1}^{-}\Delta _{2}^{-})(\Delta _{3}^{+}+\Delta _{3}^{-}).  \label{PPP}
\end{eqnarray}%
The first term is the result we expect. The middle term can be subtracted
away by means of an overall anti-Hermitian $\Omega $ correction for the
diagram. However, the last term cannot be adjusted that way, which would
require a non anti-Hermitian correction. Luckily, it disappears by itself,
as we show right away. 
\begin{figure}[t]
\begin{center}
\includegraphics[width=16truecm]{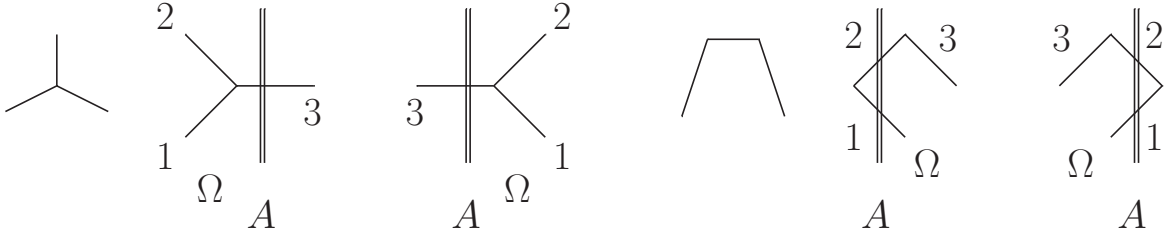}
\end{center}
\caption{Subdiagrams that need $\Omega $ correction}
\label{TripleCav}
\end{figure}

The point is that at $\Omega =0$ we miss the whole second line of formula (%
\ref{ups}). So doing, we ignore not only the overall $\Omega $ corrections
to the diagram, we can be adjusted when needed, but also the $\Omega $
corrections inherited from the subdiagrams, which cannot be neglected, nor
modified. Expanding the right-hand side in powers of $V$, formula (\ref{ups}%
) truncates to%
\begin{equation}
V_{\Omega }=V-\frac{1}{2}VBV+\frac{1}{4}VBVBV-\frac{1}{8}VBVBVBV+\Omega +%
\frac{1}{2}\Omega AV+\frac{1}{2}VA\Omega .  \label{PPPP}
\end{equation}%
The third to last term, $\Omega $, can be used to adjust the middle term of (%
\ref{PPP}): this is the overall $\Omega $ correction to the diagram. The
last two terms are the crucial ones, because they are inherited from\ the
subdiagrams. Now we show that they remove the difficulty mentioned above.

Specifically, we have to use the $\Omega $\ of formula (\ref{delta12}) for
the subdiagrams made by the two adjacent legs $\varphi _{1}$ and $\varphi
_{2}$. The last two contributions to (\ref{PPPP}) are shown to the left in
fig. \ref{TripleCav}, where the double line denotes the $A$ cut. Since $A$
is the projector onto the physical subspace $W_{\text{ph}}$, the $A$ cut can
only cross physical legs, in our case just $\varphi _{3}$. Noting that the
conventions for the orientations of the energy flows turn (\ref{delta12})
into $\Delta _{1}^{+}\Delta _{2}^{+}+\Delta _{1}^{-}\Delta _{2}^{-}$, the
last two contribution to (\ref{PPPP}) are%
\begin{equation*}
(\Delta _{1}^{+}\Delta _{2}^{+}+\Delta _{1}^{-}\Delta _{2}^{-})\Delta
_{3}^{\pm }.
\end{equation*}%
Once we include them, as per formula (\ref{PPPP}), we find the expected,
factorized result, which is%
\begin{equation*}
V_{\Omega }\big(\scalebox{1.5}{$\perp$},\text{PV}^{2}\text{-Ph}\big)=(i%
\mathcal{P}_{1})(i\mathcal{P}_{2})(i\mathcal{F}_{3}),
\end{equation*}%
again in agreement with what predicted by the diagrammatics of the theories
of physical and purely virtual particles.

Now we consider the case where $\varphi _{2}$ and $\varphi _{3}$ are both
physical, and only $\varphi _{1}$ needs to be purely virtual. At $\Omega =0$
we must drop the diagrams containing a cut that does not cross the leg $%
\varphi _{1}$. Again, the last two terms of (\ref{PPPP}) tell us that we
have to include the $\Omega $ corrections for the subdiagrams. The
interested subdiagrams are two: the one made by the legs $\varphi _{1}$ and $%
\varphi _{2}$, plus the one made by the legs $\varphi _{1}$ and $\varphi
_{3} $. Note that there is no $\Omega $ correction for the subdiagram made
by the legs $\varphi _{2}$ and $\varphi _{3}$, because $A$ cannot cut the
leg $\varphi _{1}$. We also have to include an overall $\Omega $ correction,
corresponding to the third to last term of (\ref{PPPP}), to subtract the
anti-Hermitian contributions%
\begin{equation*}
-\left[ (i\mathcal{P}_{2})(\Delta _{1}^{+}\Delta _{3}^{+}+\Delta
_{1}^{-}\Delta _{3}^{-})+(i\mathcal{P}_{3})(\Delta _{1}^{+}\Delta
_{2}^{+}+\Delta _{1}^{-}\Delta _{2}^{-})\right] .
\end{equation*}%
At the end, we find the desired, factorized result%
\begin{equation*}
V_{\Omega }\big(\scalebox{1.5}{$\perp$},\text{PV-Ph}^{2}\big)=(i\mathcal{P}%
_{1})(i\mathcal{F}_{2})(i\mathcal{F}_{3}).
\end{equation*}

\medskip

The second example of tree diagram with three legs is the one where the
propagators are adjacent, shown to the right of fig. \ref{TripleProp}. We
take%
\begin{equation*}
\mathcal{L}_{I}=K_{1}\varphi _{1}+K_{12}\varphi _{1}\varphi
_{2}+K_{23}\varphi _{2}\varphi _{3}+K_{3}\varphi _{3}.
\end{equation*}%
If all the legs are to be quantized as purely virtual, formula (\ref{ups})
gives%
\begin{equation}
(i\mathcal{P}_{1})(i\mathcal{P}_{2})(i\mathcal{P}_{3})-\left[ (i\mathcal{P}%
_{1})(\Delta _{2}^{+}\Delta _{3}^{-}+\Delta _{2}^{-}\Delta _{3}^{+})+\text{
cyclic permutations}\right]   \label{fff}
\end{equation}%
at $\Omega =0$, the energy flows being ordered according to the sequence $%
x_{1}$-$x_{2}$-$x_{3}$. Again, the terms between the square brackets, which
violate the factorization rule, can be subtracted away by means of an
overall $\Omega $ correction. No $\Omega $ corrections due to subdiagrams
contribute, since there is no physical leg that can be cut by $A$ (which is
just $|0\rangle \langle 0|$).

If the third leg is physical, the other two being purely virtual, the right
result, which is $(i\mathcal{P}_{1})(i\mathcal{P}_{2})(i\mathcal{F}_{3})$,
is obtained by including, again, the overall $\Omega $ correction that
subtracts the terms in square brackets of (\ref{fff}), plus the $\Omega $
corrections due to the subdiagram made by the two adjacent legs $\varphi _{1}
$ and $\varphi _{2}$. Here the convention for the energy flow orientations
is the same as in (\ref{delta12}).

If the middle leg is physical and the other two are purely virtual, we
obtain the desired result, which is $(i\mathcal{P}_{1})(i\mathcal{F}_{2})(i%
\mathcal{P}_{3})$, with the same overall $\Omega $ correction as for (\ref%
{fff}). No $\Omega $ corrections for subdiagrams contribute, because the
subdiagrams obtained by cutting the physical leg are just simple propagators.

If the physical legs are at the sides and the middle leg is purely virtual,
we obtain the factorized result $(i\mathcal{F}_{1})(i\mathcal{P}_{2})(i%
\mathcal{F}_{3})$ after including: $i$) the $\Omega $ corrections for the
two subdiagrams made by a physical leg and a purely virtual one, and $ii$)
the overall $\Omega $ correction, which now reads%
\begin{equation}
\left[ (i\mathcal{P}_{1})(\Delta _{2}^{+}\Delta _{3}^{-}+\Delta
_{2}^{-}\Delta _{3}^{+})+(i\mathcal{P}_{3})(\Delta _{1}^{+}\Delta
_{2}^{-}+\Delta _{1}^{-}\Delta _{2}^{+})\right] ,  \label{overall}
\end{equation}%
instead of the square bracket of (\ref{fff}).

An interesting case is when the legs $\varphi _{1}$ and $\varphi _{2}$ are
physical and the leg $\varphi _{3}$ is purely virtual. We obtain%
\begin{eqnarray}
&&(i\mathcal{F}_{1})(i\mathcal{F}_{2})(i\mathcal{P}_{3})-(\Delta
_{1}^{+}+\Delta _{1}^{-})(\Delta _{2}^{+}\Delta _{3}^{-}+\Delta
_{2}^{-}\Delta _{3}^{+})-(\Delta _{3}^{+}+\Delta _{3}^{-})(\Delta
_{1}^{+}\Delta _{2}^{-}+\Delta _{1}^{-}\Delta _{2}^{+})  \notag \\
&&-\left[ (i\mathcal{P}_{1})(\Delta _{2}^{+}\Delta _{3}^{-}+\Delta
_{2}^{-}\Delta _{3}^{+})+(i\mathcal{P}_{2})(\Delta _{1}^{+}\Delta
_{3}^{-}+\Delta _{1}^{-}\Delta _{3}^{+})\right]  \label{line}
\end{eqnarray}%
at $\Omega =0$. The first contribution, $(i\mathcal{F}_{1})(i\mathcal{F}%
_{2})(i\mathcal{P}_{3})$, is the expected, factorized result. We obtain it
after including the right $\Omega $ corrections as follows. The second line
of (\ref{line}) is subtracted by means of a new, overall $\Omega $
correction. The middle terms of the first line are subtracted by the $\Omega 
$ corrections of formula (\ref{delta12}), due to the subdiagram made by the
legs $\varphi _{2}$ and $\varphi _{3}$. The right terms of the first line
are subtracted in a new, probably unexpected way: they are canceled by the $%
\Omega $ corrections, derived in formula (\ref{O1}) below, associated with
the disconnected subdiagrams made by the leg $\varphi _{3}$ and the endpoint
of the leg $\varphi _{1}$. These corrections, illustrated in the last two
diagrams of fig. \ref{TripleCav}, read%
\begin{equation*}
\frac{1}{2}VA\Omega \rightarrow \frac{1}{2}\left( 2\Delta _{1}^{-}2\Delta
_{2}^{+}\right) \frac{\Delta _{3}^{+}+\Delta _{3}^{-}}{2},\qquad \frac{1}{2}%
\Omega AV\rightarrow \frac{1}{2}\frac{\Delta _{3}^{+}+\Delta _{3}^{-}}{2}%
\left( 2\Delta _{1}^{+}2\Delta _{2}^{-}\right) .
\end{equation*}%
The arrows stand for dropping the source factors $iK$.

We see that only when we take care of everything properly, we obtain the
desired, factorized result, and find agreement with the diagrammatics of a
theory of physical and purely virtual particles. What is important is that
we can always determine the needed $\Omega $ corrections, and that they are
unique.

\section{Disconnected diagrams}

\label{product}\setcounter{equation}{0}

Now we study the disconnected diagrams, which unexpectedly hide a number of
nontrivial caveats.

We start from the product of two constant vertices, with%
\begin{equation*}
\mathcal{L}_{I}=\lambda _{1}K_{1}+\lambda _{2}K_{2},
\end{equation*}%
the constants $\lambda _{1}$ and $\lambda _{2}$ being inserted to make the
discussion more transparent.

Nothing is propagating, so we just have the vacuum state $|0\rangle $. As an
exercise, let us first check what happens if we take $A=0$, $B=C=|0\rangle
\langle 0|=\mathbb{I}$ at $\Omega =0$. Differentiating $V_{\Omega }=V_{0}$
with respect to $iK$ for each source, and then setting the sources to zero,
we find%
\begin{eqnarray}
\left. \frac{\delta ^{2}V_{0}}{i\delta K_{1}(x)i\delta K_{2}(y)}\right\vert
_{K=0} &=&\left. \frac{\delta ^{2}V}{i\delta K_{1}(x)i\delta K_{2}(y)}%
\right\vert _{K=0}-\frac{1}{2}\left. \frac{\delta ^{2}V}{i\delta K_{1}(x)}%
\right\vert _{K=0}\left. \frac{\delta ^{2}V}{i\delta K_{2}(y)}\right\vert
_{K=0}  \notag \\
&&-\frac{1}{2}\left. \frac{\delta ^{2}V}{i\delta K_{2}(x)}\right\vert
_{K=0}\left. \frac{\delta ^{2}V}{i\delta K_{1}(y)}\right\vert _{K=0}=\lambda
_{1}\lambda _{2}-\frac{1}{2}\lambda _{1}\lambda _{2}-\frac{1}{2}\lambda
_{2}\lambda _{1}=0.\qquad 
\end{eqnarray}%
We cannot use the $\Omega $ arbitrariness to correct this result into the
expected one, $\lambda _{1}\lambda _{2}$, because $\Omega $ should be
anti-Hermitian. The reason why we find zero, instead of $\lambda _{1}\lambda
_{2}$ is that, by taking $A=0$, we have subtracted too much, including the
contributions of the vacuum state.

If we take $A=|0\rangle \langle 0|$, we have nothing to subtract ($B=0$), so
the result of formula (\ref{ups}) for $V_{\Omega }$ at $\Omega =0$ is just $%
\lambda _{1}\lambda _{2}$, i.e., the product of the two vertices.

Let us now consider the product of a propagator and a constant vertex. We
start from%
\begin{equation*}
\mathcal{L}_{I}=K_{1}\varphi _{1}+K_{2}\varphi _{1}+K_{3}.
\end{equation*}%
Formula (\ref{ups}) gives, at $A=\Omega =0$, 
\begin{equation*}
-\frac{1}{2}(\Delta _{1}^{+}+\Delta _{1}^{-}),
\end{equation*}%
which can be subtracted away by an appropriate $\Omega $ correction. Then,
the final result is zero, but, again, we have subtracted too much.

The physical space $W_{\text{ph}}$ cannot be empty: it must contain at least
the vacuum state $|0\rangle $. If we want to quantize $\varphi _{1}$ as a
purely virtual particle, we must take $A=|0\rangle \langle 0|$. Then (\ref%
{Thetasol}) gives%
\begin{equation}
i\mathcal{P}_{1}-\frac{1}{2}(\Delta _{1}^{+}+\Delta _{1}^{-}).  \label{c2}
\end{equation}%
The expected result for a purely virtual particle is not this, but just $i%
\mathcal{P}_{1}$. We obtain $i\mathcal{P}_{1}$ by means of the $\Omega $
subtraction 
\begin{equation}
\left. \frac{\delta ^{3}\Omega }{i\delta K_{1}i\delta K_{2}i\delta K_{3}}%
\right\vert _{K=0}=\frac{1}{2}(\Delta _{1}^{+}+\Delta _{1}^{-}).  \label{O1}
\end{equation}%
We see that the $\Omega $ corrections are crucial and generically
nontrivial, even in an arrangement as simple as the product of a propagator
times a constant.

The product of a purely virtual propagator and two constant vertices can be
studied by taking%
\begin{equation*}
\mathcal{L}_{I}=K_{1}\varphi _{1}+K_{2}\varphi _{1}+K_{3}+K_{4}.
\end{equation*}%
Then (\ref{Thetasol}), or (\ref{ups}), give%
\begin{equation*}
i\mathcal{P}_{1}-\Delta _{1}^{+}-\Delta _{1}^{-}
\end{equation*}%
for $A=|0\rangle \langle 0|$ at $\Omega =0$. The bad news is that we cannot
subtract the last two terms of this expression by means of an anti-Hermitian 
$\Omega $ for the overall diagram. The good news is that there is no need
to, because they disappear by themselves once we include the $\Omega $
corrections (\ref{O1}) due to the disconnected subdiagrams made by a
propagator and a single vertex, as required by the last two terms of (\ref%
{PPPP}). At the end, the result is just $i\mathcal{P}_{1}$, as desired.

The disconnected product of two purely virtual propagators is studied from%
\begin{equation*}
\mathcal{L}_{I}=K_{1}\varphi _{1}+K_{1}^{\prime }\varphi _{1}+K_{2}\varphi
_{2}+K_{2}^{\prime }\varphi _{2},
\end{equation*}%
by differentiating with respect to each $iK$ once, and then setting the
sources to zero. If we apply formula (\ref{ups}) with $A=|0\rangle \langle
0| $, $\Omega =0$, we find%
\begin{equation}
V_{0}(|~|,\text{PV}^{2})=i\mathcal{P}_{1}i\mathcal{P}_{2}-\frac{1}{2}\left[ i%
\mathcal{P}_{1}\left( \Delta _{2}^{+}+\Delta _{2}^{-}\right) +i\mathcal{P}%
_{2}\left( \Delta _{1}^{+}+\Delta _{1}^{-}\right) \right] .  \label{twoprop}
\end{equation}%
We obtain the expected result, $V_{\Omega }(|~|,$PV$^{2})=i\mathcal{P}_{1}i%
\mathcal{P}_{2}$, once we remove the terms in square brackets by means of an
overall $\Omega $ correction.

If the leg $\varphi _{1}$ is purely virtual and the leg $\varphi _{2}$ is
physical, formula (\ref{Thetasol}) gives%
\begin{equation}
V_{0}(|~|,\text{PV-Ph})=i\mathcal{P}_{1}i\mathcal{F}_{2}-\left( \Delta
_{1}^{+}+\Delta _{1}^{-}\right) \left( \Delta _{2}^{+}+\Delta
_{2}^{-}\right) -\frac{1}{2}\left[ i\mathcal{P}_{2}\left( \Delta
_{1}^{+}+\Delta _{1}^{-}\right) \right] .  \label{twoprop2}
\end{equation}%
The middle term is subtracted by including the $\Omega $ correction (\ref{O1}%
) associated with the disconnected subdiagrams made by the $\varphi _{1}$
propagator and any endpoint of the $\varphi _{2}$ propagator. The last term
is subtracted by the overall $\Omega $ correction. At the end, we find the
expected, factorized result $V_{\Omega }(|~|,$PV-Ph$)=i\mathcal{P}_{1}i%
\mathcal{F}_{2}$.

We have learned that the projections of the tree diagrams and those of the
disconnected diagrams are not as straightforward as we might have hoped.
Yet, they always give the expected, factorized results once we choose the $%
\Omega $ corrections appropriately. The examples we have studied suggest
that the right $\Omega $ is determined uniquely by this requirement.

\section{One-loop diagrams}

\label{oneloop}\setcounter{equation}{0}

In this section and the next one we study loop diagrams. We use the
conventions of \cite{diagrammarMio}. Specifically, we integrate on the loop
energies $k^{0}$, with measure $\mathrm{d}k^{0}/(2\pi )$, and ignore the
integrals on the space components $\mathbf{k}$ of the loop momenta. The
reason is that the identities we write hold for arbitrary values of the
frequencies $\omega $ of the internal and external legs of the diagrams.
Moreover, we conventionally multiply every propagator by a factor $2\omega $%
. So doing, we obtain the so-called \textquotedblleft skeleton
diagrams\textquotedblright , which allow us to study unitarity by means of
simple algebraic operations.

Every internal leg is labeled by an index $a,b,\ldots $. The $a$-th leg has
mass $m_{a}$ and carries momentum $k^{\mu }-p_{a}^{\mu }$, where $k^{\mu
}=(k^{0},\mathbf{k})$ denotes the loop momentum and $p_{a}^{\mu }=(e_{a},%
\mathbf{p}_{a})$ is an external momentum. The frequency of the $a$-th leg is 
$\omega _{a}=\sqrt{m_{a}^{2}+(\mathbf{k}-\mathbf{p}_{a})^{2}}$. In the
notation we are adopting, each internal leg has its own external momentum $%
p_{a}$. So doing, the external momenta are redundant, but make the formulas
more symmetric and easier to handle.

After multiplying by $2\omega $, the Feynman propagator $P$, its conjugate $%
P^{\ast }$ and the cut propagators $P^{\pm }$ become%
\begin{equation}
P=\frac{i}{e-\omega +i\epsilon }-\frac{i}{e+\omega -i\epsilon },\qquad
P^{\ast }=\frac{i}{e+\omega +i\epsilon }-\frac{i}{e-\omega -i\epsilon }%
,\qquad P^{\pm }=(2\pi )\delta (e\mp \omega ).  \label{propaga}
\end{equation}

For example, the skeleton of a one-loop Feynman diagram with $N$ internal
legs is%
\begin{equation}
G_{N}^{s}=\int \frac{\mathrm{d}k^{0}}{2\pi }\prod\limits_{a=1}^{N}\left( 
\frac{i}{k^{0}-e_{a}-\omega _{a}+i\epsilon _{a}}-\frac{i}{k^{0}-e_{a}+\omega
_{a}-i\epsilon _{a}}\right) .  \label{GsN}
\end{equation}%
We have a different overall factor with respect to \cite{diagrammarMio},
since we assume that the vertices are equal to one (while in \cite%
{diagrammarMio} they are equal to $-i$).

For future use, we define%
\begin{eqnarray*}
\Delta ^{ab} &=&\pi \delta (e_{a}-e_{b}-\omega _{a}-\omega _{b}),\qquad 
\mathcal{P}^{ab}=\mathcal{P}\frac{1}{e_{a}-e_{b}-\omega _{a}-\omega _{b}}, \\
\mathcal{Q}^{ab} &=&\mathcal{P}^{ab}-\mathcal{P}\frac{1}{e_{a}-e_{b}-\omega
_{a}+\omega _{b}},\qquad \qquad \mathcal{\hat{P}}^{ab}=\mathcal{P}^{ab}+%
\mathcal{P}^{ba}.
\end{eqnarray*}%
In particular, $\mathcal{P}^{ab}$ and $\Delta ^{ab}$ are the basic
ingredients of the \textquotedblleft threshold
decomposition\textquotedblright\ of a skeleton diagram $G^{s}$, which
organizes the contributions to $G^{s}$ according to the number of delta
functions, which are on shell. This number is called \textit{level} of the
decomposition.

We recall that, starting from the threshold decomposition of a Feynman
diagram, the threshold decomposition of a diagram with purely virtual
particles f is obtained by suppressing the delta functions whose arguments
contain any frequency $\omega _{\text{f}}$ of such particles \cite%
{diagrammarMio}.

In all the examples we consider, $C=\mathbb{I}$, $A$ is the projector $\Pi _{%
\text{ph}}$ onto the physical space $W_{\text{ph}}$, and $B=\mathbb{I}-\Pi _{%
\text{ph}}$ is the projector onto the complement $W_{\text{pv}}$. We study
the case where all the internal legs are quantized as purely virtual
particles ($A=|0\rangle \langle 0|$), as well as the cases where some
internal legs are physical and the others are purely virtual.

\subsection{Bubble and double bubble}

To study the bubble diagram, we take%
\begin{equation*}
\mathcal{L}_{I}=K\varphi _{1}\varphi _{2}+K^{\prime }\varphi _{1}\varphi _{2}
\end{equation*}%
differentiate with respect to $iK$ and $iK^{\prime }$ and then set $%
K=K^{\prime }=0$. After integrating on the loop energy, the skeleton bubble
diagram with two physical internal legs is 
\begin{equation}
i\mathcal{\hat{P}}^{12}+\Delta ^{12}+\Delta ^{21},  \label{Feybu}
\end{equation}%
as in \cite{diagrammarMio}, apart from the different overall sign, due to
the new notation for the vertices.

With two purely virtual internal legs, or one physical leg and one purely
virtual leg, formula (\ref{Thetasol}), or formula (\ref{ups}) at $\Omega =0$%
, give the skeleton 
\begin{equation}
G_{2}^{s}\big(\text{PV}^{2}\big)=i\mathcal{\hat{P}}^{12}.  \label{PV2}
\end{equation}%
This result is the right one for the purely virtual bubble \cite%
{diagrammarMio}, so there in no need of overall $\Omega $ corrections. There
are no $\Omega $ corrections inherited from subdiagrams.

The square bubble (two bubble diagrams with a common vertex) is studied by
taking%
\begin{equation*}
\mathcal{L}_{I}=K\varphi _{1}\varphi _{2}+K^{\prime }\varphi _{1}\varphi
_{2}\varphi _{3}\varphi _{4}+K^{\prime \prime }\varphi _{3}\varphi _{4}
\end{equation*}%
and following the usual procedure. First, we quantize all the internal legs
as purely virtual. Formula (\ref{ups}) gives%
\begin{equation}
(i\mathcal{\hat{P}}^{12})(i\mathcal{\hat{P}}^{34})-\Delta ^{12}\Delta
^{43}-\Delta ^{21}\Delta ^{34}  \label{proj}
\end{equation}%
at $\Omega =0$, where the energies $e_{1}$ and $e_{3}$ flow from the left to
the right, while the energies $e_{2}$ and $e_{4}$ flow from the right to the
left. This is not the result of \cite{diagrammarMio}: the last two terms of (%
\ref{proj}) should not be there.

The point is that the double bubble is a product of diagrams, in momentum
space. As before, the projection (\ref{proj}) of the product does not
coincide with the product $(i\mathcal{\hat{P}}^{12})(i\mathcal{\hat{P}}%
^{34}) $ of the projected factors, at $\Omega =0$. Yet, it is sufficient to
choose an anti-Hermitian $\Omega $ such that 
\begin{equation}
\left. \frac{\delta ^{3}\Omega }{i\delta Ki\delta K^{\prime }i\delta
K^{\prime \prime }}\right\vert _{K=0}=\Delta ^{12}\Delta ^{43}+\Delta
^{21}\Delta ^{34}  \label{Omdoububble}
\end{equation}%
to subtract away the last two terms of (\ref{proj}). In the end, $V_{\Omega
} $ gives the expected result, as in \cite{diagrammarMio}.

If the first bubble contains one or two purely virtual legs and the second
bubble contains two physical legs, formula (\ref{ups}) at $\Omega =0$ gives%
\begin{equation*}
(i\mathcal{\hat{P}}^{12})(i\mathcal{\hat{P}}^{34}+\Delta ^{34}+\Delta
^{43})-\Delta ^{12}\Delta ^{43}-\Delta ^{21}\Delta ^{34}.
\end{equation*}%
The last two terms are subtracted, again, by means of the $\Omega $
correction (\ref{Omdoububble}). In the end, we obtain the expected result,
i.e, the product of a purely virtual bubble times a physical bubble.

\subsection{Triangle}

\label{tr}

Now we study the triangle diagram. We take%
\begin{equation*}
\mathcal{L}_{I}=K_{12}\varphi _{1}\varphi _{2}+K_{23}\varphi _{2}\varphi
_{3}+K_{31}\varphi _{3}\varphi _{1}.
\end{equation*}%
differentiate with respect to $iK$ once for every $K$ and then set $K=0$. We
recall that the threshold decomposition of the triangle made of Feynman
propagators is \cite{diagrammarMio}%
\begin{equation}
G_{3}^{s}=-\mathcal{P}_{\text{3}}+i\sum_{\text{perms}}\Delta ^{ab}\mathcal{Q}%
^{ac}+\frac{1}{2}\sum_{\text{perms}}\Delta ^{ab}(\Delta ^{ac}+\Delta ^{cb}),
\label{GS3}
\end{equation}%
where 
\begin{equation*}
\mathcal{P}_{\text{3}}=\frac{1}{2}\sum_{\text{perms}}(\mathcal{P}^{ab}%
\mathcal{P}^{ac}+\mathcal{P}^{ba}\mathcal{P}^{ca})=\mathcal{P}^{12}\mathcal{P%
}^{13}+\text{cycl}+(e\rightarrow -e)
\end{equation*}%
is its purely virtual part. An extra factor $i$ for every vertex with
respect to \cite{diagrammarMio} is due to the different notation we are
using here for the vertices.

\begin{figure}[t]
\begin{center}
\includegraphics[width=16truecm]{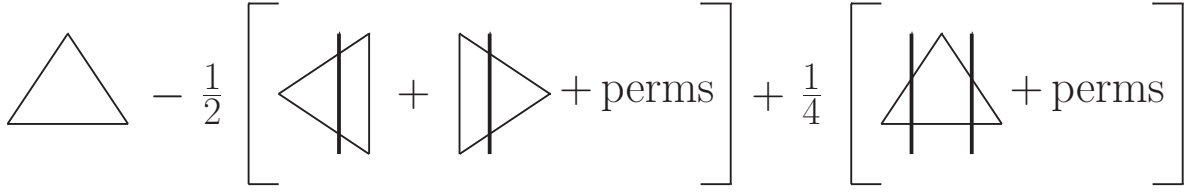}
\end{center}
\caption{{}Triangle}
\label{Triangle}
\end{figure}

We first study the case where all the internal legs have to be quantized as
purely virtual. The reduced amplitude $V_{\Omega }$ of formula (\ref{ups})
at $\Omega =0$ gives the diagrams shown in fig. \ref{Triangle}. The result
is the same as in \cite{diagrammarMio}, i.e.,%
\begin{equation}
G_{3}^{s}\big(\text{PV}^{3}\big)=-\mathcal{P}_{\text{3}}.  \label{PVtriangle}
\end{equation}%
This means that we do not need overall $\Omega $ corrections. Moreover,
there are no $\Omega $ corrections due to subdiagrams.

For completeness, we report the two basic cut diagrams of fig. \ref{Triangle}%
:%
\begin{equation}
2i\Delta ^{32}[\mathcal{Q}^{31}-i\Delta ^{31}-i\Delta ^{12}],\qquad \qquad
\qquad 4\Delta ^{13}\Delta ^{23}.  \label{diagra}
\end{equation}%
The former is the triangle with a single cut, where the uncut leg 1 is
placed on the right-hand side. The latter is the triangle with two cuts,
where leg 3 is cut twice and the vertex $\varphi _{1}\varphi _{3}$ is placed
on the left-hand side. The other diagrams are obtained from (\ref{diagra})
by means of permutations, or by flipping the signs of the energies. 

If one internal leg is physical and the other two have to be quantized as
purely virtual, the result is the same, because all the diagrams of fig. \ref%
{Triangle} still contribute. Instead, if two internal legs ($\varphi _{2}$
and $\varphi _{3}$) are physical and the other one is purely virtual, we
must drop the diagrams that have no field $\varphi _{1}$, or two fields $%
\varphi _{1}$, to the left or right of a $B$ cut, as explained in section %
\ref{PVphysics}. We obtain the result%
\begin{equation}
G_{3}^{s}\big(\text{PV-Ph}^{2}\big)=-\mathcal{P}_{\text{3}}+i\mathcal{Q}%
^{21}\Delta ^{23}+i\mathcal{Q}^{31}\Delta ^{32},  \label{GS3-12}
\end{equation}%
for $\Omega =0$, which agrees again with the one of \cite{diagrammarMio}.

We see that we never need $\Omega $ corrections for triangle diagrams.

\subsection{Box}

\label{qu}

The box diagram is studied from 
\begin{equation*}
\mathcal{L}_{I}=K_{12}\varphi _{1}\varphi _{2}+K_{23}\varphi _{2}\varphi
_{3}+K_{34}\varphi _{3}\varphi _{4}+K_{41}\varphi _{4}\varphi _{1},
\end{equation*}%
with the usual procedure. The threshold decomposition of the skeleton made
of Feynman propagators, derived in ref. \cite{diagrammarMio}, reads 
\begin{equation}
G_{4}^{s}\,=-i\mathcal{P}_{4}-\frac{1}{2}\sum_{\text{perms}}\Delta ^{ab}%
\mathcal{Q}^{ac}\mathcal{Q}^{ad}+\frac{i}{2}\sum_{\text{perms}}\Delta
^{ab}(\Delta ^{ac}+\Delta ^{cb})\mathcal{Q}^{ad}+\frac{1}{6}\sum_{\text{perms%
}}\Delta ^{ab}(\Delta ^{ac}\Delta ^{ad}+\Delta ^{cb}\Delta ^{db}),
\label{box}
\end{equation}%
where%
\begin{equation}
-i\mathcal{P}_{4}\equiv -\frac{i}{6}\sum_{\text{perms}}\mathcal{P}^{ab}%
\mathcal{P}^{ac}\mathcal{P}^{ad}-\frac{i}{4}\sum_{\text{perms}}\mathcal{P}%
^{ab}\mathcal{P}^{ac}\mathcal{P}^{db}+(e\rightarrow -e)  \label{P4}
\end{equation}%
is the purely virtual part of the diagram.

Again, we find that formula (\ref{ups}) with $A=|0\rangle \langle 0|$ gives 
\begin{equation}
G_{4}^{s}\big(\text{PV}^{4}\big)=-i\mathcal{P}_{4}  \label{GS4}
\end{equation}%
at $\Omega =0$, with matches the result of \cite{diagrammarMio}, when all
the internal legs are purely virtual. When one internal leg is physical and
the other three are purely virtual, the result is the same.

\begin{figure}[t]
\begin{center}
\includegraphics[width=14truecm]{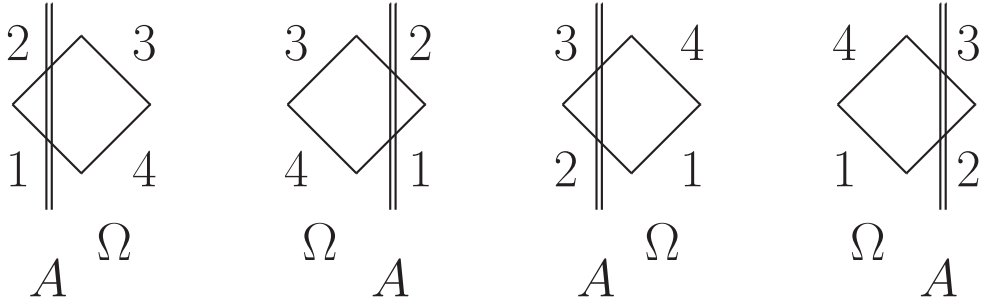}
\end{center}
\caption{$\Omega $ corrections from box subdiagrams}
\label{Box}
\end{figure}

When two adjacent internal legs (say, $\varphi _{1}$ and $\varphi _{2}$) are
physical and the other two must be quantized as purely virtual, we obtain%
\begin{equation*}
-i\mathcal{P}_{4}-\Delta ^{12}\mathcal{Q}^{13}\mathcal{Q}^{14}-\Delta ^{21}%
\mathcal{Q}^{23}\mathcal{Q}^{24}-\left[ \Delta ^{12}(\Delta ^{32}\Delta
^{14}+\Delta ^{42}\Delta ^{13})+\Delta ^{21}(\Delta ^{31}\Delta ^{24}+\Delta
^{41}\Delta ^{23})\right]
\end{equation*}%
from $V_{\Omega }$ at $\Omega =0$. The result of \cite{diagrammarMio} is
made by the first three terms of this expression. The final part of the
formula, the one in square brackets, cannot be subtracted away by means of
an overall $\Omega $ correction for the diagram. Luckily, it disappears
automatically when we include, as per the last two terms of (\ref{PPPP}),
the $\Omega $ corrections (\ref{delta12}) due to the subdiagrams made by the
two purely virtual legs, shown in the first two drawings of fig. \ref{Box}.
We easily find%
\begin{equation*}
\frac{1}{2}VA\Omega \rightarrow \Delta ^{21}(\Delta ^{31}\Delta ^{24}+\Delta
^{41}\Delta ^{23}),\qquad \qquad \frac{1}{2}\Omega AV\rightarrow \Delta
^{12}(\Delta ^{13}\Delta ^{42}+\Delta ^{14}\Delta ^{32}),
\end{equation*}%
and finally get%
\begin{equation}
G_{4}^{s}\big(\text{Ph}^{2}\text{-PV}^{2}\big)=-i\mathcal{P}_{4}-\Delta ^{12}%
\mathcal{Q}^{13}\mathcal{Q}^{14}-\Delta ^{21}\mathcal{Q}^{23}\mathcal{Q}%
^{24}.  \label{GS4-22}
\end{equation}%

When the legs $\varphi _{1}$ and $\varphi _{3}$ are physical, while the legs 
$\varphi _{2}$ and $\varphi _{4}$ are purely virtual, formula (\ref{ups}) at 
$\Omega =0$ gives%
\begin{equation}
G_{4}^{s}\big(\text{Ph-PV-Ph-PV}\big)=-i\mathcal{P}_{4}-\Delta ^{13}\mathcal{%
Q}^{12}\mathcal{Q}^{14}-\Delta ^{31}\mathcal{Q}^{32}\mathcal{Q}^{34},
\label{PhPV4}
\end{equation}%
which agrees with the result of \cite{diagrammarMio}. In this case, no $%
\Omega $ corrections are involved, since the subdiagrams obtained by cutting
the physical legs are just simple propagators.

Finally, when three internal legs are physical and only $\varphi _{4}$ is
purely virtual, $V_{\Omega }$ gives 
\begin{eqnarray}
&&-i\mathcal{P}_{4}-\sum_{p(1,2,3)}\Delta ^{ab}\mathcal{Q}^{ac}\mathcal{Q}%
^{a4}+\frac{i}{2}\sum_{p(1,2,3)}\Delta ^{ab}(\Delta ^{ac}+\Delta ^{cb})%
\mathcal{Q}^{a4}  \notag \\
&&\qquad \qquad -\left[ \Delta ^{12}(\Delta ^{32}\Delta ^{14}+\Delta
^{42}\Delta ^{13})+\Delta ^{21}(\Delta ^{31}\Delta ^{24}+\Delta ^{41}\Delta
^{23})\right]   \notag \\
&&\qquad \qquad -\left[ \Delta ^{14}(\Delta ^{34}\Delta ^{12}+\Delta
^{24}\Delta ^{13})+\Delta ^{41}(\Delta ^{31}\Delta ^{42}+\Delta ^{21}\Delta
^{43})\right]   \label{GS4-31}
\end{eqnarray}%
at $\Omega =0$, where $p(1,2,3)$ is the set of permutations $\{a,b,c\}$ of
1, 2 and 3. The result of \cite{diagrammarMio} is just the first line. The
other two lines are subtracted by the $\Omega $ corrections (\ref{delta12})
that originate from the subdiagrams made by the legs $\varphi _{3}$ and $%
\varphi _{4}$, and the subdiagrams made by the legs $\varphi _{1}$ and $%
\varphi _{4}$, respectively, shown in fig. \ref{Box}. As before, no $\Omega $
corrections come from the $A$ cut of the legs $\varphi _{1}$ and $\varphi
_{3}$. The final result, given by the complete $V_{\Omega }$ formula, is thus%
\begin{equation}
G_{4}^{s}\big(\text{Ph}^{3}\text{-PV}\big)=-i\mathcal{P}_{4}-\sum_{p(1,2,3)}%
\Delta ^{ab}\mathcal{Q}^{ac}\mathcal{Q}^{a4}+\frac{i}{2}\sum_{p(1,2,3)}%
\Delta ^{ab}(\Delta ^{ac}+\Delta ^{cb})\mathcal{Q}^{a4},  \label{GS4-31fin}
\end{equation}%
as desired.

\section{Diagrams with more loops}

\label{moreloop}\setcounter{equation}{0}

\begin{figure}[t]
\begin{center}
\includegraphics[width=12truecm]{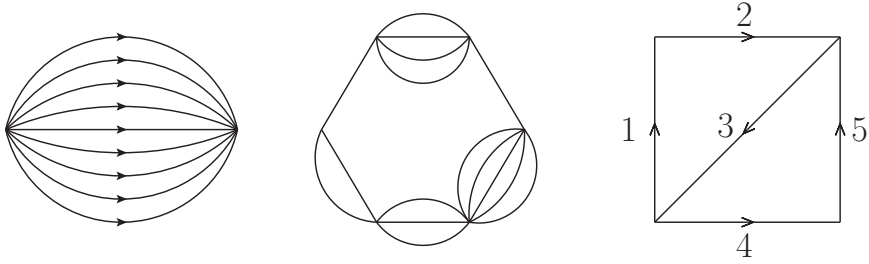}
\end{center}
\caption{Stacks of propagators and box diagram with diagonal}
\label{stack}
\end{figure}

In this section we study diagrams with more loops. The diagrams of a certian
subclass are equivalent to one-loop diagrams, and can be treated with no
extra effort. This happens when an internal leg is replaced by a stack of
legs with the same endpoints, as shown to the left of fig. \ref{stack}. At
the level of skeleton diagrams, we have the identity%
\begin{equation}
\mathcal{S}_{n}^{s}=\mathcal{S}_{1}^{s}\left(
\sum_{i=1}^{n}e_{i},\sum_{i=1}^{n}\omega _{i}\right) ,  \label{ideF}
\end{equation}%
where $\mathcal{S}_{n}^{s}$ denotes the skeleton of the diagram made by the
stack of $n$ propagators, $\mathcal{S}_{1}^{s}$ is a single propagator, $%
e_{i}$ are the external energies of the various legs, oriented from right to
left, and $\omega _{i}$ are frequencies of the legs. The identity (\ref{ideF}%
) holds for \textquotedblleft Feynman stacks\textquotedblright , as well as
\textquotedblleft non time-ordered stacks\textquotedblright . The former are
made by $n$ Feynman propagators, in which case $\mathcal{S}_{1}^{s}$ is a
single Feynman propagator $P$ of (\ref{propaga}). The latter are made by $n$
cut propagators, in which case $S_{1}^{s}$ is a single cut propagator $P^{+}$
of (\ref{propaga}). 

If the stack $\mathcal{S}_{n}^{s}$ contains one or more purely virtual legs,
while the other legs are Feynman propagators, no overall $\Omega $
correction is required, as well as no $\Omega $ corrections for subdiagrams.
Combining the facts just stated, the reduction $V\rightarrow V_{0}$ of
formula (\ref{Thetasol}) shows that the projection of the stack $\mathcal{S}%
_{n}^{s}$ is just the propagator%
\begin{equation}
\mathcal{P}\left( \frac{i}{\sum_{i=1}^{n}(e_{i}-\omega _{i})}-\frac{i}{%
\sum_{i=1}^{n}(e_{i}+\omega _{i})}\right)  \label{stackpro}
\end{equation}%
of a single purely virtual particle with energy equal to the total incoming
energy and frequency equal to the total frequency.

The bubble with \textquotedblleft pseudodiagonal\textquotedblright\ is the
stack $n=3$. If all the legs are physical, its expression is the analogue of
(\ref{Feybu}), i.e., (\ref{ideF}) for a Feynman stack. If a leg is purely
virtual, the reduced amplitude is (\ref{stackpro}).

The triangle with pseudodiagonal\ is the triangle where one leg is replaced
by a stack $\mathcal{S}_{2}^{s}$. Let us assume that the first and fourth
legs have the same endpoints, and their energies $e_{1}$ and $e_{4}$ have
the same orientations. Then, we easily retrieve the formulas of subsection %
\ref{tr} with $e_{1}\rightarrow e_{1}+e_{4}$, $\omega _{1}\rightarrow \omega
_{1}+\omega _{4}$.

The same works for the box with pseudodiagonal, where the endpoints of the
fifth leg coincide with those of one of the first four legs. And the same
works for diagrams with arbitrarily many loops: when the internal legs can
be grouped together into stacks, the results coincide with those of a
diagram with fewer loops, obtained by replacing each stack with a single
leg, with energy equal to the total energy flowing into the stack, and
frequency equal to the total frequency. An example is shown in the middle of
fig. \ref{stack}, which is a diagram equivalent to the hexagon. These
properties hold for the Feynman diagrams, as well as for the diagrams of the
reduced scattering matrices derived in section \ref{key}.

The first nontrivial arrangement at two loops is the box with (true)
diagonal, shown to the right of fig. \ref{stack}. The arrows are opposite to
the orientations of the external energies $e_{i}$. From \cite{diagrammarMio}%
, the threshold decomposition reads%
\begin{eqnarray}
G_{4D}^{s} &=&-i\mathcal{P}_{4D}+\Delta ^{12}\tilde{G}_{3|345}^{s\text{PV}%
}+\Delta ^{21}\tilde{G}_{3|345}^{s\text{PV}}+\Delta ^{45}\tilde{G}_{3|123}^{s%
\text{PV}}+\Delta ^{54}\tilde{G}_{3|123}^{s\text{PV}}  \notag \\
&&+\frac{i}{2}\sum_{s4D}(\mathcal{Q}^{a3c}-2i\Delta ^{a3c})\Delta
^{ab}\Delta ^{cd}+i\sum_{s4D}\Delta ^{a3c}\left[ i\mathcal{Q}^{ab}(\mathcal{Q%
}^{cd}-i\Delta ^{cd})+\Delta ^{ab}\mathcal{Q}^{cd}\right]  \notag \\
&&+\frac{i}{2}\sum_{s4D}\Delta ^{a3c}\left[ \Delta ^{a3d}(\mathcal{Q}%
^{ab}-i\Delta ^{ab})+\Delta ^{b3c}(\mathcal{Q}^{cd}-i\Delta ^{cd})\right] ,
\label{Box4D}
\end{eqnarray}%
where%
\begin{equation*}
\mathcal{P}_{4D}=\sum_{s4D}\mathcal{P}^{a3c}\left[ \mathcal{P}^{ab}\mathcal{P%
}^{cd}+\frac{1}{2}\mathcal{P}^{ab}\mathcal{P}^{a3d}+\frac{1}{2}\mathcal{P}%
^{b3c}\mathcal{P}^{cd}+\frac{1}{2}\mathcal{P}^{a3d}\mathcal{P}^{b3c}\right] ,
\end{equation*}%
and $\Delta ^{ab}\tilde{G}_{3|cdf}^{s\text{PV}}=\Delta ^{ab}\left.
G_{3|cdf}^{s\text{PV}}\right\vert _{e_{3}\rightarrow e_{3}-e_{b}-\omega
_{b}} $, $G_{3|abc}^{s\text{PV}}$ being the skeleton (\ref{PVtriangle}) of
the purely virtual triangle with legs $abc$. Moreover, $\mathcal{P}^{a3c}$, $%
\mathcal{Q}^{a3c}$ and $\Delta ^{a3c}$ are the same as $\mathcal{P}^{a3}$, $%
\mathcal{Q}^{a3}$ and $\Delta ^{a3}$, respectively, with $e_{a}\rightarrow
e_{a}+e_{c}$ and $\omega _{a}\rightarrow \omega _{a}+\omega _{c}$. The sums $%
\sum_{s4D}$ are over the permutations $a,b\ $of $1,2$, the permutations $%
c,d\ $of $4,5$, plus $(e\rightarrow -e)$.

We start from the case where every internal leg is quantized as purely
virtual. If we apply formula (\ref{ups}) with $\Omega =0$, we find%
\begin{equation}
-i\mathcal{P}_{4D}-i\left[ \Delta ^{21}\Delta ^{54}\mathcal{Q}%
^{235}+(e\rightarrow -e)\right]  \label{pad}
\end{equation}%
which is the expected result, $-i\mathcal{P}_{4D}$, plus a term that can be
canceled by means of an overall anti-Hermitian $\Omega $ correction. In the
end, we obtain 
\begin{equation}
G_{4D}^{s}\big(\text{PV}^{5}\big)=-i\mathcal{P}_{4D},  \label{Gs4D}
\end{equation}%
as desired.

The reason why it is necessary to include the correction just mentioned is
easily explained. Although $G_{4D}^{s}$ is a prime diagram, it factorizes
when we \textquotedblleft contract\textquotedblright\ the diagonal. The
contraction operation, denoted by $C_{3}$, is studied in detail in section %
\ref{purevirtuality}. It amounts to multiplying the skeleton diagram by $%
im_{3}/2$ and taking the limit $m_{3}\rightarrow \infty $.

Diagrammatically, the result of the contraction is a purely virtual double
bubble, which is obviously not prime. We know from section \ref{oneloop}
that such a diagram needs the $\Omega $ correction (\ref{Omdoububble}). A
consistency check is to multiply the square bracket of (\ref{pad}) by $%
im_{3}/2$, take the limit $m_{3}\rightarrow \infty $, and verify that what
we obtain is indeed canceled by the analogue of (\ref{Omdoububble}).

Now we consider the physical box with a purely virtual diagonal, i.e., the
case where the diagonal is the only purely virtual leg. This time, the
contraction $C_{3}$ gives the physical double bubble, which does not need
any $\Omega $ correction. Indeed, formula (\ref{ups}) at $\Omega =0$
correctly gives%
\begin{equation*}
G_{4D}^{s}\big(\text{Ph}^{2}\text{-PV-Ph}^{2}\big)=\left.
G_{4D}^{s}\right\vert _{3},
\end{equation*}%
where $\left. G_{4D}^{s}\right\vert _{a,b,\cdots }$ means the expression (%
\ref{Box4D}) upon suppression of all the terms where any frequency $\omega
_{a}$, $\omega _{b}$, $\cdots $ appears in the argument of some delta
function. No $\Omega $ corrections for subdiagrams are involved, since the
subdiagrams in question are triangles.

When the legs 1 and 3 are purely virtual, while all the other ones are
physical, the contraction $C_{3}$ gives, again, the purely virtual double
bubble, so the same $\Omega $ correction as for (\ref{pad}) must be
included. The reduced amplitude $V_{\Omega }$ gives%
\begin{equation*}
G_{4D}^{s}\big(\text{PV-Ph-PV-Ph}^{2}\big)=\left. G_{4D}^{s}\right\vert
_{1,3},
\end{equation*}%
in agreement with \cite{diagrammarMio}.

An interesting case is the one where two adjacent non diagonal legs, say 2
and 5, are purely virtual, while the others are physical. We find the same
overall $\Omega $ correction as above, because $C_{3}$ gives the purely
virtual double bubble. However, we also find some unwanted terms that cannot
be subtracted by means of an overall, anti-Hermitian $\Omega $. Luckily,
they cancel out automatically, as in the other cases analyzed so far, once
we include the $\Omega $ corrections due to the subdiagrams, as per the last
two terms $\Omega AV/2$ and $VA\Omega /2$ of formula (\ref{PPPP}).

The unwanted terms are subtracted away by the diagrams where the three
physical legs, which are 1, 3 and 4, are crossed by the $A$ cut. One side of
the cut contains the vertex $\varphi _{1}\varphi _{3}\varphi _{4}$, while
the other side contains the $\Omega $ corrections (\ref{delta12}) to the
subdiagram made by the legs 2 and 5. At the end, we correctly find 
\begin{equation*}
G_{4D}^{s}\big(\text{Ph-PV-Ph}^{2}\text{-PV}\big)=\left.
G_{4D}^{s}\right\vert _{2,5},
\end{equation*}

The other cases can be treated similarly.

\section{Managing pure virtuality: methods and theorems}

\label{purevirtuality}\setcounter{equation}{0}

In this section and the next one we describe methods and tricks to study the
virtual and on-shell contents of the skeleton diagrams $G^{s}$, and relate
the threshold decompositions of different diagrams to one another. We can
even derive the threshold decompositions of bigger diagrams from the ones of
smaller diagrams in a unique way. The results allow us to gain insight into
the threshold decompositions themselves and the roles of the $\Omega $
corrections. In section \ref{rules} we recap the lessons learned through the
various examples.

The two main tricks are integration and contraction, which stand for: $a$)
integrating on the external energies, and $b$) sending the masses to
infinity.

\subsection{Integration}

Using the notation (\ref{propaga}) (where, we recall, we multiply every
propagator by $2\omega $ with respect to the usual definitions), a basic
tool is to integrate a skeleton diagram $G^{s}$ on an independent external
energy $e$:%
\begin{equation}
I_{e}(G^{s})\equiv \int_{-\infty }^{+\infty }\frac{\mathrm{d}e}{2\pi }%
G^{s}(e).  \label{PVinte}
\end{equation}%
The virtue of this operation is that it turns the Feynman propagator, as
well as the cut propagators into unity:%
\begin{equation}
\int_{-\infty }^{+\infty }\frac{\mathrm{d}e}{2\pi }\left( \frac{i}{e-\omega
+i\epsilon }-\frac{i}{e+\omega -i\epsilon }\right) =\int_{-\infty }^{+\infty
}\frac{\mathrm{d}e}{2\pi }(2\pi )\delta (e\pm \omega )=1.  \label{ava}
\end{equation}%
Moreover, it turns a purely virtual (tree) propagator into zero:%
\begin{equation}
\mathcal{P}\int_{-\infty }^{+\infty }\frac{\mathrm{d}e}{2\pi }\left( \frac{i%
}{e-\omega }-\frac{i}{e+\omega }\right) =0.  \label{avv}
\end{equation}

We denote the operation (\ref{PVinte}), applied to the internal leg $\ell $,
by $I_{\ell }$, or $I_{e_{\ell }}$. It is a useful tool to inspect the
diagrams and, for example, check whether they are purely virtual or not, and
quantify their virtual contents versus their on-shell contents. It is also
useful, as we show in the next section, to ascend and descend among the
diagrams.

If we prefer to use the standard notation (\ref{prop}), (\ref{cutprop}),
then the operation $I_{\ell }$ is%
\begin{equation*}
I_{e}(\bar{G}^{s})\equiv 2\omega \int_{-\infty }^{+\infty }\frac{\mathrm{d}e%
}{2\pi }\bar{G}^{s}(e),
\end{equation*}%
where $\bar{G}^{s}$ denotes the skeleton diagram with propagators (\ref{prop}%
), (\ref{cutprop}).

We can apply $I_{\ell }$ to one or more internal legs of a diagram $G^{s}$.
We begin by studying what happens when we integrate on all the independent
external energies $e_{i}$. Taking into account that, by definition, we are
already integrating on all the internal energies of a skeleton diagram, we
end up integrating on all the independent energies of the diagram. So doing,
all the Feynman propagators collapse to unity.

The result of this operation is called \textit{avirtuality} of the skeleton
diagram $G^{s}$ and measures its \textquotedblleft
on-shellness\textquotedblright . If $G^{s}$ is the skeleton of an ordinary
Feynman diagram with $v$ (nonderivative) vertices, its avirtuality is equal
to one\footnote{%
We recall that we are working in the notation where each nonderivative
vertex is equal to 1. In the usual notation, where a vertex is equal to $%
-i\lambda $, $\lambda $ denoting some coupling, we would have $(-i\lambda
)^{v}$. Diagrams with derivative vertices can be reduced to sums of diagrams
with nonderivative vertices, as explained in \cite{diagrammarMio}.}.

The purely virtual contents of the one-loop prime diagrams considered so far
were singled out by\ reduced amplitude $\mathring{V}$ of formula (\ref%
{primeT}). Denote the skeleton diagrams associated with $\mathring{V}$ by $%
\mathring{G}^{s}$. We show that the avirtuality of an arbitrary $\mathring{G}%
^{s}$ with $v$ vertices is%
\begin{equation}
\mathrm{A}_{v}=1+\sum_{k=1}^{v-1}\frac{(-1)^{k}}{2^{k}}a_{k+1,v},
\label{avirtuality}
\end{equation}%
where $a_{k+1,v}$ is defined by the recursive relation 
\begin{equation}
a_{n,v}=n^{v}-\sum_{k=1}^{n-1}\binom{n}{k}a_{k,v},\qquad a_{1,v}=1.
\label{recurra}
\end{equation}

Formula (\ref{avirtuality}) is proved as follows. Expand the right-hand side
of equation (\ref{primeT}) in powers of $v$. The $n$th power contains $n-1$
vertical cuts, and each cut is equal to 
\begin{equation*}
-\frac{1}{2}(C-|0\rangle \langle 0|).
\end{equation*}%
The cuts identify $n-2$ vertical strips. Two half planes lie at the sides,
their boundaries being the first and the last cuts. We consider them as
further strips. The $v$ vertices of $\mathring{G}^{s}$ must be distributed
inside the $n$ strips in all possible ways. If we include the possibility to
leave some strips empty, there are $n^{V}$ ways of doing so. However, such a
possibility must be excluded.

Let $a_{n,v}$ denote the number of distributions where such a possibility is
indeed excluded. Clearly, $a_{n,v}$ is equal to $n^{v}$ minus the
distributions that contain empty strips. Such distributions can be
distinguished according to the number $k$ of empty strips, which ranges from 
$1$ to $n-1$. There are $\binom{n}{k}$ ways of choosing the $k$ empty
strips. In each case, the $v$ vertices can be distributed in $a_{n-k,v}$
ways. This gives the recurrence relation (\ref{recurra}).

Each arrangement gives a contribution equal to unity, once the operation (%
\ref{PVinte}) is applied to all the independent external energies. Expanding
(\ref{primeT}), we thus find formula (\ref{avirtuality}).

We see that the avirtuality of a diagram depends only on the number of
vertices, not on the type of diagram $\mathring{G}^{s}$. We can easily check
that $\mathrm{A}_{v}$ vanishes for every even $v$. The avirtualities of the
first odd values of $v$ are%
\begin{equation*}
\mathrm{A}_{1}=1,\qquad \mathrm{A}_{3}=-\frac{1}{2},\qquad \mathrm{A}%
_{5}=1,\qquad \mathrm{A}_{7}=-\frac{17}{4},\qquad \mathrm{A}_{9}=31,\qquad 
\mathrm{A}_{11}=-\frac{691}{2}.
\end{equation*}

Let us apply formula (\ref{avirtuality}) to the examples of the previous
sections. The simplest case is the avirtuality of the purely virtual
propagator (\ref{PVprop}), which we know to be zero by formula (\ref{avv}).
This is the case $v=2$.

\subsubsection*{Triangle}

Now we check that the avirtuality of the triangle of purely virtual
particles is indeed $\mathrm{A}_{3}=-1/2$, using formula (\ref{PVtriangle}).
To this purpose, we need the identity (\ref{idainte}), proved in appendix %
\ref{app2}. We have to calculate 
\begin{equation*}
I_{e_{1}}I_{e_{2}}\hspace{0.02in}G_{3}^{s}\big(\text{PV}^{3}\big)%
=-\int_{-\infty }^{+\infty }\frac{\mathrm{d}e_{1}}{2\pi }\int_{-\infty
}^{+\infty }\frac{\mathrm{d}e_{2}}{2\pi }\left[ \mathcal{P}^{12}\mathcal{P}%
^{13}+\text{ cycl }+(e\rightarrow -e)\right] .
\end{equation*}%
Particular attention has to be paid to the convergence of the integrals at
infinity. Ignoring the integral over $e_{1}$ for a moment, we find%
\begin{equation}
-\int_{-\infty }^{+\infty }\frac{\mathrm{d}e_{2}}{2\pi }(\mathcal{P}^{23}%
\mathcal{P}^{21}+\mathcal{P}^{32}\mathcal{P}^{12})-\int_{-\infty }^{+\infty }%
\frac{\mathrm{d}e_{2}}{2\pi }(\mathcal{P}^{12}+\mathcal{P}^{23})\mathcal{P}%
^{13}-\int_{-\infty }^{+\infty }\frac{\mathrm{d}e_{2}}{2\pi }(\mathcal{P}%
^{21}+\mathcal{P}^{32})\mathcal{P}^{31}.  \label{sepa}
\end{equation}%
The last two integrals of this list are convergent for $e_{2}\rightarrow \pm
\infty $, and give zero. The first integral can be worked out by means of (%
\ref{idainte}) and gives%
\begin{equation}
I_{e_{2}}\hspace{0.02in}G_{3}^{s}\big(\text{PV}^{3}\big)=-\frac{\pi }{2}%
[\delta (e_{1}-e_{3}-\omega _{1}+\omega _{3})+\delta (e_{1}-e_{3}+\omega
_{1}-\omega _{3})].  \label{triV}
\end{equation}%
At this point, the operation $I_{e_{1}}$ gives $-1/2$, as we wanted to show.

One may wonder if we can add $\Omega $ corrections to obtain zero, instead.
If not, we must infer that (\ref{triV}) and $\mathrm{A}_{3}=-1/2$ are
intrinsic to the diagram.

We recall that $\Omega $ cannot change the level 0 of the threshold
decomposition, which matches the Euclidean diagram. It cannot change the odd
levels of the decomposition either, because they are not anti-Hermitian. So,
the first level that can be affected by $\Omega $ is the second one. There
are two possibilities to remove (\ref{triV}) by means of $\Omega $. One is
to add something containing $\Delta ^{12}\Delta ^{32}$, plus $(e\rightarrow
-e)$, because the operation $I_{e_{2}}$ on it can compensate (\ref{triV}).
However, the symmetries under the permutations of the internal legs, and $%
(e\rightarrow -e)$, imply that we would have to add the whole sum 
\begin{equation*}
\Delta ^{12}\Delta ^{32}+\text{cycl }+(e\rightarrow -e).
\end{equation*}%
The operation $I_{e_{2}}$ on the additional terms gives contributions
proportional to $\Delta ^{13}$, which must not be there.

The second possibility is to add%
\begin{equation*}
\pi ^{2}[\delta (e_{1}-e_{3}-\omega _{1}+\omega _{3})\delta
(e_{1}-e_{2}-\omega _{1}+\omega _{2})+\delta (e_{1}-e_{3}+\omega _{1}-\omega
_{3})\delta (e_{1}-e_{2}+\omega _{1}-\omega _{2})].
\end{equation*}%
This is not acceptable either, since the double singularities due to these
delta functions are not present in the starting triangle diagram. A quick
way to see this is by noting that differences of frequencies appear, which
cannot be traded for sums of frequencies. By stability, it must be possible
to express all the singularities of a skeleton diagram defined by means of
the Feynman $i\epsilon $ prescription in terms of sums of frequencies (see 
\cite{diagrammarMio} for details).

We conclude that the avirtuality $\mathrm{A}_{3}=-1/2$ is an intrinsic
property of the purely virtual triangle diagram.

\subsubsection*{Box}

The avirtuality $\mathrm{A}_{4}$ of the box diagram with circulating purely
virtual particles is equal to zero. We can verify this result as before,
from formulas (\ref{P4}) and (\ref{GS4}), using the identities (\ref{idainte}%
) and (\ref{ide3}) of appendix \ref{app2}.

We apply $I_{e_{3}}I_{e_{2}}I_{e_{1}}$ to (\ref{GS4}): we first integrate on 
$e_{1}$, then on $e_{2}$ and finally on $e_{3}$. We can distinguish terms $%
\mathcal{P}^{ab}\mathcal{P}^{cd}\mathcal{P}^{ef}$ with three, two and one
indices equal to 1. The $e_{1}$ integral gives zero on the terms that just
have one index 1, since they can be organized into the sums%
\begin{equation*}
-\frac{i}{2}\int_{-\infty }^{+\infty }\frac{\mathrm{d}e_{1}}{2\pi }\mathcal{P%
}^{ab}\mathcal{P}^{ac}\left( 2\mathcal{P}^{a1}+\mathcal{P}^{1b}+\mathcal{P}%
^{1c}\right) =-\frac{i}{2}\int_{-\infty }^{+\infty }\frac{\mathrm{d}e_{1}}{%
2\pi }\mathcal{P}^{ba}\mathcal{P}^{ca}\left( 2\mathcal{P}^{1a}+\mathcal{P}%
^{b1}+\mathcal{P}^{c1}\right) =0,
\end{equation*}%
where $\{a,b,c\}$ is any permutation of $\{2,3,4\}$. These integrals are
separately convergent.

Each term with two or three indices equal to 1 is separately convergent, and
can be calculated by means of (\ref{idainte}) and (\ref{ide3}). After the
operation $I_{e_{2}}$, the result is%
\begin{equation*}
I_{e_{2}}I_{e_{1}}G_{4}^{s}\big(\text{PV}^{4}\big)=-\frac{i}{2}(\mathcal{P}%
^{34}+\mathcal{P}^{43}).
\end{equation*}%
When we finally apply $I_{e_{3}}$, we get zero, as we wanted to show.

\medskip

Following the same guidelines, we have checked the avirtualities of the
purely virtual pentagon and the purely virtual hexagon (formulas (\ref%
{pentha}), $\mathrm{A}_{5}=1$, $\mathrm{A}_{6}=0$), finding agreement with (%
\ref{avirtuality}). The projection $\mathring{V}$ of the purely virtual box
diagram with diagonal gives (\ref{pad}), which also satisfies $\mathrm{A}%
_{4}=0$. The inclusion of the $\Omega $ correction, which leads to (\ref%
{Gs4D}), does not change $\mathrm{A}_{4}$.

\subsection{Contraction}

Another useful operation is the limit of infinite masses, after multiplying
by the masses themselves. The basic identities are%
\begin{equation}
\lim_{m^{2}\rightarrow \infty }\frac{(im^{2})i}{p^{2}-m^{2}\pm i\epsilon }%
=1,\qquad \lim_{m^{2}\rightarrow \infty }(im^{2})(2\pi )\theta (\pm
p^{0})\delta (p^{2}-m^{2})=0,  \label{basic}
\end{equation}%
which select the principal-value part and kill the on-shell part, and allow
us to measure the \textit{virtuality} of a skeleton diagram. The operation,
which we denote by $C_{\ell }$, where the letter $C$ stands for
\textquotedblleft contraction\textquotedblright\ and the suffix $\ell $
denotes the leg that is being contracted, has other interesting virtues. The
first one is that it allows us to jump from one diagram to a simpler
diagram, in momentum space. Later on we show that it also allows us to jump
from simpler diagrams to more complicated diagrams, once it is combined with
the integration trick mentioned before.

We start from an ordinary Feynman skeleton diagram $G^{s}$, and denote the
diagram obtained by contracting the leg $\ell $ by $C_{\ell }(G^{s})$. For
example, if $G^{s}$ is the box and $\ell $ is one of its internal legs, $%
C_{\ell }(G^{s})$ is the triangle. If $G^{s}$ is the triangle, $C_{\ell
}(G^{s})$ the bubble. If $G^{s}$ is the bubble, $C_{\ell }(G^{s})$ is the
tadpole. If we contract the third leg of the diagram made by three adjacent
propagators, we obtain the diagram made by two adjacent propagators. Etc.
Note that a connected diagram $G^{s}$ is mapped into a connected diagram $%
C_{\ell }(G^{s})$. Instead, a prime diagram $G^{s}$ can be mapped into a
factorized diagram $C_{\ell }(G^{s})$. For example, we have seen that the
box with diagonal turns into the double bubble, by contraction of the
diagonal leg.

The other interesting property of the operation $C_{\ell }$ is that it
applies straightforwardly to the projected skeleton diagrams, the reduced
scattering matrices, and every term of the expansion of the right-hand side
of formula (\ref{ups}), including the $\Omega $ corrections. Precisely: the
contraction $C_{\ell }$ and the projection $V_{\Omega }$ commute.

To prove this statement, we work on the amplitude $V_{\Omega }$ of (\ref{ups}%
), starting from $\Omega =0$. Let $\hat{G}^{s}$ denote the projection of the
Feynman skeleton diagram $G^{s}$, $\widehat{C_{\ell }(G^{s})}$ the
projection of the contracted diagram $C_{\ell }(G^{s})$, and $C_{\ell }(\hat{%
G}^{s})$ the contraction of the projected diagram $\hat{G}^{s}$. By (\ref%
{basic}), $C_{\ell }$ sends the cut propagators of the leg $\ell $ to zero.
This means that every cut diagram contributing to $\hat{G}^{s}$, where the
leg $\ell $ is crossed by a cut, disappears. Thus, the surviving diagrams of 
$C_{\ell }(\hat{G}^{s})$ are precisely the ones of $\widehat{C_{\ell }(G^{s})%
}$. We conclude that the projection encoded in the reduced amplitude $%
V_{\Omega }$ commutes with the contraction $C_{\ell }$ at $\Omega =0$: $%
\widehat{C_{\ell }(G^{s})}=C_{\ell }(\hat{G}^{s})$.

It then follows that the two operations also commute at nonzero $\Omega $,
if $\Omega $ is the one that gives a theory of physical and purely virtual
particles. The reason is that the diagrammatics of a theory of physical and
purely virtual particles, recalled at the beginning of section \ref{oneloop}%
, amounts to a projection that manifestly commutes with the contraction:
starting from the threshold decomposition of a Feynman diagram, it
suppresses the delta functions whose arguments contain the frequencies $%
\omega _{\text{f}}$ of the purely virtual particles f. The contraction $%
C_{\ell }$ suppresses the delta functions that contain $\omega _{\ell }$,
via the second limit of (\ref{basic}). The order with which we remove the
two is clearly immaterial.

\medskip

These $C_{\ell }$ properties can be used to relate the $\Omega $ corrections
of bigger skeleton diagrams to the $\Omega $ corrections of smaller
diagrams, and check the results of the previous sections. For example, if we
apply $C_{2}$ to (\ref{resdo}), we find the purely virtual propagator (\ref%
{PVprop}), with $m\rightarrow m_{1}$. If we apply $C_{2}$ to (\ref{delta12})
we find 0, since the single propagator has no $\Omega $. If we apply $C_{3}$
to (\ref{resdokid}), we find (\ref{resdo}) (with $\Delta _{2}^{\pm
}\rightarrow \Delta _{2}^{\mp }$, due to the different conventions for the
energy flows). If we apply $C_{3}$ to the difference between (\ref{fffres})
and (\ref{resdokid}), we correctly find (\ref{delta12}) (again with $\Delta
_{2}^{\pm }\rightarrow \Delta _{2}^{\mp }$). If we apply $C_{3}$ to (\ref%
{fff}) we obtain (\ref{resdo}) again. If we apply $C_{3}$ to (\ref{line}),
we obtain $i\mathcal{F}_{1}i\mathcal{F}_{2}$, as expected. If we apply $%
C_{2} $ to (\ref{line}), we correctly obtain (\ref{resdoph}) with $%
2\rightarrow 3$. If we apply $C_{2}$ to (\ref{twoprop}), we find (\ref{c2}).
If we apply $C_{2}$ to (\ref{twoprop2}), we also find (\ref{c2}). If we
apply $C_{1}$ to (\ref{twoprop2}), we correctly find the physical propagator 
$i\mathcal{F}_{2}$. And so on.

In loop diagrams we switch to the notation where every propagator is
multiplied by $2\omega $ with respect to the usual definitions. The basic
identities are then%
\begin{equation}
\lim_{m\rightarrow \infty }\frac{im}{2}\frac{i}{e\pm (\omega -i\epsilon )}%
=\mp \frac{1}{2},\qquad \lim_{m\rightarrow \infty }\frac{im}{2}\mathcal{P}%
\frac{i}{e\pm \omega }=\mp \frac{1}{2},\qquad \lim_{m\rightarrow \infty }%
\frac{im}{2}(2\pi )\delta (e\pm \omega )=0.  \notag
\end{equation}%
It is easy to check that $C_{4}$ turns the purely virtual box skeleton (\ref%
{GS4}) into the purely virtual triangle skeleton (\ref{PVtriangle}), and $%
C_{3}$ turns (\ref{PVtriangle}) into the purely virtual bubble skeleton (\ref%
{PV2}). Similarly, $C_{4}$ turns (\ref{PhPV4}) into (\ref{GS3-12}) with $%
1\leftrightarrow 2$, and $C_{4}$ turns (\ref{GS4-31fin}) into (\ref{GS3}),
etc.

The operation $C_{\ell }$ preserves the threshold decomposition, by which we
mean that it maps level $i$ to level $i$, for each $i$. For example, the
decomposition (\ref{box}) of the box diagram is sent term by term into the
decomposition (\ref{GS3}) of the triangle diagram, by the operation $C_{4}$.
The integration operation $I_{\ell }$ considered before, instead, mixes
different levels (see below).

\section{Ascending and descending among skeleton diagrams}

\label{ascension}\setcounter{equation}{0}

Now we explain how to use the operations $I_{\ell }$ and $C_{\ell }$ to
ascend and descend among the skeleton diagrams. We have already described
the descent $C_{\ell }$ in various cases, which is rather straightforward
and preserves the threshold decomposition. The operation $I_{\ell }$,
instead, deserves a more detailed analysis.

We start from purely virtual diagrams. They are simpler, because they just
contain principal values $\mathcal{P}^{ab}$, and no delta functions $\Delta
^{ab}$. We have already proved the descent relations%
\begin{equation*}
G_{4}^{s}\big(\text{PV}^{4}\big)\overset{C_{4}}{\longrightarrow }G_{3}^{s}%
\big(\text{PV}^{3}\big)\overset{C_{3}}{\longrightarrow }G_{2}^{s}\big(\text{%
PV}^{2}\big)\overset{C_{2}}{\longrightarrow }G_{1}^{s}\big(\text{PV}\big)=0,
\end{equation*}%
where $G_{1}^{s}\big($PV$\big)$ is the purely virtual tadpole, which
vanishes. Using the formulas of \cite{diagrammarMio}, we can extend these
results to the hexagon and the pentagon:%
\begin{equation*}
G_{6}^{s}\big(\text{PV}^{6}\big)\overset{C_{6}}{\longrightarrow }G_{5}^{s}%
\big(\text{PV}^{5}\big)\overset{C_{5}}{\longrightarrow }G_{4}^{s}\big(\text{%
PV}^{4}\big),
\end{equation*}%
where\footnote{%
We point out two typos in \cite{diagrammarMio}: the factors $1/5!$ and $1/6!$
in front of the pentagon and hexagon expressions reported there, formula
(8.2), should be replaced by $1/4!$ and $1/5!$, respectively, to match the
analogous factors of the triangle and the box. Further factors $i$ and $-1$
in formulas (\ref{pentha}) are due to the different notation we are using
here for the vertices.}%
\begin{align}
G_{5}^{s}\big(\text{PV}^{5}\big)& =\frac{1}{4!}\sum_{\text{perms}}\mathcal{P}%
^{ab}\mathcal{P}^{ac}\left[ \mathcal{P}^{ad}(\mathcal{P}^{ae}+4\mathcal{P}%
^{ed})+2\mathcal{P}^{db}\mathcal{P}^{ec}\right] +(e_{i}\rightarrow -e_{i}), 
\notag \\
G_{6}^{s}\big(\text{PV}^{6}\big)& =\frac{i}{5!}\sum_{\text{perms}}\mathcal{P}%
^{ab}\mathcal{P}^{ac}\mathcal{P}^{ad}\left[ \mathcal{P}^{ae}(\mathcal{P}%
^{af}+5\mathcal{P}^{fe})+5\mathcal{P}^{ed}\left( \mathcal{P}^{ef}+2\mathcal{P%
}^{fc}\right) \right] +(e_{i}\rightarrow -e_{i}).  \label{pentha}
\end{align}

We want to show that we can ascend through these skeletons by means of the
sole operations $C_{\ell }$, and the requirement of correct behaviors at
large energies (which are the convergence conditions for the operations $%
I_{\ell }$).

\subsubsection*{From bubble to triangle}

\vskip -.2truecm

The purely virtual triangle can only be proportional to $\mathcal{P}_{\text{3%
}}$, because of the symmetries under the exchanges of the internal legs, and 
$e\rightarrow -e$. The proportionality constant can be fixed from the purely
virtual bubble (\ref{PV2}), by requiring $G_{3}^{s}\big($PV$^{3}\big)\overset%
{C_{3}}{\longrightarrow }G_{2}^{s}\big($PV$^{2}\big)$. We then find (\ref%
{PVtriangle}).

\subsubsection*{From triangle to box}

\vskip -.2truecm

By the symmetries mentioned above, the purely virtual box can only be a
linear combination%
\begin{equation}
-a_{1}\frac{i}{6}\sum_{\text{perms}}\mathcal{P}^{ab}\mathcal{P}^{ac}\mathcal{%
P}^{ad}-a_{2}\frac{i}{4}\sum_{\text{perms}}\mathcal{P}^{ab}\mathcal{P}^{ac}%
\mathcal{P}^{db}+(e\rightarrow -e).  \label{puta4}
\end{equation}%
It is obtained by listing the monomials $\mathcal{P}^{a_{1}a_{2}}\mathcal{P}%
^{a_{3}a_{4}}\cdots $ according to the following rules (codified by the
\textquotedblleft snowflake diagrams\textquotedblright\ of ref. \cite%
{diagrammarMio}): each index $a_{i}$ must appear at least once; if it is
repeated, it must be always to the left, or always to the right; no squares
or higher powers of the same $\mathcal{P}^{a_{i}a_{j}}$ can appear; the
monomial $\mathcal{P}^{a_{1}a_{2}}\mathcal{P}^{a_{3}a_{4}}\cdots $ cannot
factorize into the product of unlinked monomials (which means: monomial
factors with no index in common).

Next, the total should be integrable in every independent external energy.
In particular, it should decrease faster than $1/e_{4}$ for large $e_{4}$.
Applied to (\ref{puta4}), this condition gives $a_{2}=a_{1}$. It is easy to
check that the requirement $G_{4}^{s}\big($PV$^{4}\big)\overset{C_{4}}{%
\longrightarrow }G_{3}^{s}\big($PV$^{3}\big)$ then implies $a_{1}=1$, thus
giving (\ref{GS4}).

\subsubsection*{From box to pentagon}

\vskip -.2truecm

The purely virtual pentagon must be a linear combination%
\begin{equation*}
\sum_{\text{perms}}\mathcal{P}^{ab}\mathcal{P}^{ac}\left[ \mathcal{P}%
^{ad}(a_{1}\mathcal{P}^{ae}+a_{2}\mathcal{P}^{ed})+\mathcal{P}^{db}(a_{3}%
\mathcal{P}^{de}+a_{4}\mathcal{P}^{eb}+a_{5}\mathcal{P}^{ec})\right]
+(e_{i}\rightarrow -e_{i}),
\end{equation*}%
obtained by listing the terms with 4, 3 and 2 identical left indices. After $%
(e_{i}\rightarrow -e_{i})$, the second and forth terms are identical, as
well as the third and fifth terms, so we can set $a_{4}=a_{5}=0$. The
requirement that the total falls faster than $1/e_{5}$ for large $e_{5}$
gives $a_{2}=4a_{1}$. Finally, the requirement $G_{5}^{s}\big($PV$^{5}\big)%
\overset{C_{5}}{\longrightarrow }G_{4}^{s}\big($PV$^{4}\big)$ gives $%
a_{1}=1/4!$ and $a_{3}=1/12$. At the end, we get $G_{5}^{s}\big($PV$^{5}%
\big)
$, as in (\ref{pentha}).

\subsubsection*{From pentagon to hexagon}

\vskip -.2truecm

Listing the terms as before, the purely virtual hexagon must be a linear
combination%
\begin{equation*}
\sum_{\text{perms}}\mathcal{P}^{ab}\mathcal{P}^{ac}\left\{ \mathcal{P}^{ad}%
\left[ \mathcal{P}^{ae}(a_{1}\mathcal{P}^{af}+a_{2}\mathcal{P}^{fe})+%
\mathcal{P}^{ed}(a_{3}\mathcal{P}^{fd}+a_{4}\mathcal{P}^{ef}+a_{5}\mathcal{P}%
^{fc})\right] +a_{6}\mathcal{P}^{db}\mathcal{P}^{ec}\mathcal{P}^{ef}\right\}
,
\end{equation*}%
plus $(e_{i}\rightarrow -e_{i})$. The requirement that the total falls
faster than $1/e_{6}$ for large $e_{6}$ gives $a_{2}=5a_{1}$, $%
a_{3}=(15a_{1}+a_{4}-2a_{5})/4$ and $a_{6}=(2a_{4}-a_{5})/2$. The
requirement $G_{6}^{s}\big($PV$^{6}\big)\overset{C_{6}}{\longrightarrow }%
G_{5}^{s}\big($PV$^{5}\big)$ gives $a_{1}=i/5!$ and $a_{4}=i/4!$. The
parameter $a_{5}$ multiplies a combination that is identically zero (see 
\cite{diagrammarMio}). Setting $a_{5}=i/12$, we get the correct $G_{6}^{s}%
\big($PV$^{6}\big)$, as in (\ref{pentha}).

\subsection{Ascending through the threshold decompositions}

Similarly, we can derive the threshold decompositions of bigger skeleton
diagrams $G_{\text{big}}^{s}$ from those of smaller skeleton diagrams $G_{%
\text{small}}^{s}$. The goal is achieved by first parametrizing the most
general decompositions of $G_{\text{big}}^{s}$. After that, the arbitrary
coefficients are determined by descending to smaller diagrams $G_{\text{small%
}}^{s}$ in all possible ways by means of the operations $I_{\ell }$ and $%
C_{\ell }$. The result is unique.

We illustrate this property by studying the chain 
\begin{equation*}
\text{bubble}\rightarrow \text{triangle}\rightarrow \text{box}
\end{equation*}%
on Feynman skeletons, which means that we assume that all the internal legs
are physical. So doing, we cover all the situations obtained by the various
projections, with arbitrary combinations of physical and purely virtual
internal legs.

We have already shown how to ascend through the purely virtual versions of
the diagrams, which is equivalent to ascend through the zeroth levels of the
threshold decompositions of the Feynman diagrams. The next task is to ascend
through the other levels. We do so by writing the most general linear
combinations of the allowed terms, built with $\mathcal{P}^{ab}$ and $\Delta
^{cd}$ according to the rules explained above and satisfying the symmetries
given earlier. Then we fix the free constants by descending with the help of
the operations $I_{\ell }$ and $C_{\ell }$. The operations $C_{\ell }$
relate the decompositions level by level, so there is no need to rearrange
the decompositions after applying them. Instead, the operations $I_{\ell }$
mix different levels. This means that, after applying $I_{\ell }$ to a
bigger skeleton $G_{\text{big}}^{s}$, the result must be decomposed anew
before comparing it to the threshold decomposition of the smaller skeleton $%
G_{\text{small}}^{s}$.

\subsubsection*{From bubble to triangle}

\vskip -.2truecm

We know that the zeroth level of the threshold decomposition of the triangle
skeleton diagram is $-\mathcal{P}_{\text{3}}$. We can parametrize the most
general nonzero levels as%
\begin{eqnarray*}
\text{level 1}\text{: } &&i\alpha \sum_{\text{perms}}\Delta ^{ab}(\mathcal{P}%
^{ac}+\mathcal{P}^{cb}), \\
\text{level 2}\text{: } &&\frac{\beta }{2}\sum_{\text{perms}}\Delta
^{ab}(\Delta ^{ac}+\Delta ^{cb}).
\end{eqnarray*}%
where $\alpha $ and $\beta $ are coefficients to be determined.

First, we require that the contraction $C_{3}$ gives the bubble diagram (\ref%
{Feybu}). This implies $\alpha =1$. Then, we require that the bubble diagram
is also obtained by applying the integration $I_{3}$. This gives $\beta =1$.
At the end, we obtain the decomposition $G_{3}^{s}$ of formula (\ref{GS3}).

\subsubsection*{From triangle to box}

\vskip -.2truecm

The zeroth level of the threshold decomposition of the box skeleton diagram
was determined from the parametrization (\ref{puta4}). It coincides with $-i%
\mathcal{P}_{4}$, given in formula (\ref{P4}). Distributing the repeated
indices in all possible ways, and using the symmetries mentioned earlier,
the most general nonzero levels of the decomposition can be parametrized as%
\begin{eqnarray*}
&&\text{level 1}\text{: }\sum_{\text{perms}}\mathcal{P}^{ab}\left[ \mathcal{P%
}^{ac}(\alpha _{1}\Delta ^{ad}+\alpha _{2}\Delta ^{db})+(\alpha _{3}\mathcal{%
P}^{db}+\alpha _{4}\mathcal{P}^{dc})\Delta ^{ac}\right] +(e\rightarrow -e),
\\
&&\text{level 2}\text{: }\sum_{\text{perms}}\Delta ^{ab}\left[ \Delta
^{ac}(\beta _{1}\mathcal{P}^{ad}+\beta _{2}\mathcal{P}^{db})+(\beta
_{3}\Delta ^{db}+\beta _{4}\Delta ^{dc})\mathcal{P}^{ac}\right]
+(e\rightarrow -e), \\
&&\text{level 3}\text{: }\sum_{\text{perms}}\Delta ^{ab}\Delta ^{ac}(\gamma
_{1}\Delta ^{ad}+\gamma _{2}\Delta ^{db})+(e\rightarrow -e).
\end{eqnarray*}%
where $\alpha _{i}$, $\beta _{i}$ and $\gamma _{i}$ are the coefficients
that must be determined.

It is easy to check that $\alpha _{2}$ and $\alpha _{3}$, as well as $\beta
_{2}$ and $\beta _{3}$, multiply identical terms, so we can set $\alpha
_{3}=\beta _{3}=0$. Next, we impose the convergence of the operations $%
I_{\ell }$. This gives $\alpha _{4}=\alpha _{1}-(\alpha _{2}/2)$ and $\beta
_{2}=\beta _{1}$. Third, we require that the contraction $C_{4}$ gives the
threshold decomposition of the triangle skeleton. Matching the various
levels, we find $\alpha _{1}=-1/2$ and $\beta _{1}=i/2$.

Fourth, we require that the integration $I_{4}$ also gives the triangle.
When we apply the operation $I_{4}$, we find that it does not preserve the
levels of the threshold decomposition. The easiest way to see this is that $%
I_{4}$ returns terms that cannot be written by means of $\mathcal{P}^{ac}$
and $\Delta ^{cd}$, because they depend on differences $\omega _{i}-\omega
_{j}$ of frequencies rather than just sums $\omega _{i}+\omega _{j}$. The
corresponding singularities must cancel out, since they do not belong to the
triangle skeleton. Their cancellation is achieved by means of identities
like (\ref{identa}), whose right-hand sides contain remnants that correct
the lower levels. Once we reorganize the decomposition properly, we can
match the various levels as required. We then find $\gamma _{1}=1/6$ and $%
\gamma _{2}=-\alpha _{2}/2$.

After these substitutions we find that $\alpha _{2}$ and $\beta _{4}$
multiply trivial terms, so we can set $\alpha _{2}=\beta _{4}=0$. At the
end, we obtain the correct decomposition (\ref{box}).

\section{Diagrammatic rules recap}

\label{rules}

It is useful to summarize here the various diagrammatic options we have.

\medskip

$a$) \textbf{Feynman diagrams}

They give the usual scattering amplitudes, collected in the matrix $V=iT$.
The ingredients are the vertices and the propagators defined by the Feynman $%
i\epsilon $ prescription. The rules to build the diagrams follow from the
time-ordered product (\ref{upsop}).

\medskip

$b$) \textbf{Cutkosky-Veltman diagrams}

They are not used for the scattering amplitudes of the theory, but to
express the unitarity equation (\ref{finalized}) as a set of diagrammatic
identities. The ingredients are the same as in ($a$), plus: the conjugate
vertices, the conjugate propagators, the cut propagators. The rules to build
the diagrams follow from the time-ordered product (\ref{upsop}) and the
optical identity (\ref{finalized}). Precisely: the cut is unique; one side
of the cut is built with the rules ($a$); the other side is built with the
conjugate rules; the cut is given by cut propagators.

\medskip

$c$)\ \textbf{Minimally non time-ordered diagrams}

They give the scattering amplitudes of the reduced matrix $V_{\Omega }$. The
ingredients are the same as in ($a$), plus the non time-ordered propagators,
which coincide with the cut propagators of ($b$). No conjugate vertices, nor
conjugate propagators are involved. The instructions to assemble the
diagrams are encoded in formula (\ref{ups}). The diagrams may contain
arbitrary numbers of cuts. The rules ($a$) are used in between two cuts, and
at the sides, while the cut propagators are non time-ordered. The
anti-Hermitian matrix $\Omega $ is determined step by step to match the
amplitude $V_{\Omega }=iT_{\text{ph}}$ of a theory of physical and purely
virtual particles.

\medskip

$d$) \textbf{Cut diagrams of minimally non time-ordered diagrams}

They include the minimally non time-ordered diagrams of $V_{\Omega }$, as in
($c$), their conjugates $V{}_{\Omega }^{\dagger }$, and the gluing of $%
V{}_{\Omega }^{\dagger }$ to $V_{\Omega }$ by means of further cuts. They
are used to convert the unitarity equation (\ref{Theta}) satisfied by the
reduced amplitude $V_{\text{red}}=V_{\Omega }$ of ($c$) into diagrammatic
identities.

\section{Symmetries and renormalizability}

\label{Symmetries}

In this section we show that the projection $V\rightarrow V_{\Omega }$
preserves the symmetries of the theory, as well as its renormalizability,
under extremely mild assumptions, which are satisfied by the minimally non
time-ordered product and the new quantization principle.

The diagrammatic formulation of ref. \cite{diagrammarMio} provides a
particular solution to the problem considered here, i.e., map the usual
amplitude $V$, which is defined by the time-ordered product (\ref{upsop})
and satisfies the (pseudo)unitarity equation (\ref{C}), into a unitary
amplitude $V_{\text{ph}}=iT_{\text{ph}}$. Since (\ref{ups}) is the most
general solution to the problem, there must exist an $\Omega $ that turns $%
V_{\Omega }$ into the diagrammatics of \cite{diagrammarMio}. We denote it by 
$\Omega _{\text{ph}}(A)$. In the previous sections, we have shown how to
derive $\Omega _{\text{ph}}(A)$, by requiring that the projection of a
product diagram equals the product of its projected prime factors, and that
this factorization property survives the basic operations of ascent and
descent through the diagrams. This ensures, in particular, that $\Omega _{%
\text{ph}}(A)$ itself is diagrammatic. Its diagrams can be obtained by
comparing those of \cite{diagrammarMio} with those of $V_{0}$, encoded in
formula (\ref{Thetasol}). The comparison must be done iteratively for each
contribution to $\Omega $, as soon as it appears as an overall correction to
some diagram.

Now we show how to obtain the threshold decomposition of a diagram from
formula (\ref{ups}). Once we choose the physical space $W_{\text{ph}}$, we
know $A=\Pi _{\text{ph}}$, which is the projector onto $W_{\text{ph}}$. Let $%
V_{\text{ph}}(A)$ denote the physical solution $V_{\Omega _{\text{ph}}(A)}$.
Since $V_{\text{ph}}(A)$ depends just on $V$ ($A$ and $B=C-A$ being given),
we can invert its expression and expand $V$ in terms of $V_{\text{ph}}(A)$.

Let us do this in the particular case $A=|0\rangle \langle 0|$, where $W_{%
\text{ph}}$ is just made of the vacuum state $|0\rangle $. Then, the
projection $V_{\text{ph}}(|0\rangle \langle 0|)$ singles out the purely
virtual contents of the diagrams. Inverting the formula (\ref{ups}) of $V_{%
\text{ph}}(|0\rangle \langle 0|)$, we can write $V$ as an expansion in
powers of $V_{\text{ph}}(|0\rangle \langle 0|)$:\ this is precisely the
threshold decomposition of the Feynman diagrams collected in $V$.

The levels of the decomposition are the numbers of cuts, plus the levels of
the $\Omega _{\text{ph}}(|0\rangle \langle 0|)$ corrections. The latter are
determined by comparison with what they correct, which is easier to do when
they appear as overall corrections.

\medskip

\textbf{Lorentz invariance}

The threshold decomposition of ref. \cite{diagrammarMio} is not manifestly
Lorentz invariant, because the skeleton diagrams are defined by ignoring the
integrals on the space components of the loop momenta. It is easy to show
that Lorentz invariance is recovered when those integrals are resumed.

We recall that the projection to purely virtual particles amounts to
consider the threshold decomposition, and remove the contributions where
some delta functions, such as $\Delta ^{ab}$, depend on the frequencies $%
\omega _{\text{f}}$ of some purely virtual particles f. This operation is
Lorentz invariant (as long as it is performed consistently in all the
diagrams of the theory), since it amounts to remove a certain type of
singularity everywhere from Feynman diagrams. Different types of
singularities do not talk to one another.

Formula (\ref{ups}) is manifestly Lorentz invariant, as long as the subspace 
$W_{\text{ph}}$ is Lorentz invariant and the $\Omega $ correction is $\Omega
_{\text{ph}}(A)$. The Lorentz invariance of $\Omega _{\text{ph}}(A)$ follows
by comparison between the projection obtained here and the one of \cite%
{diagrammarMio}.

\medskip

\textbf{Gauge symmetry, general covariance, generalized local symmetries}

The physical amplitude $V_{\text{ph}}(A)$ is manifestly invariant under such
symmetries, as long as $A$ projects onto an invariant subspace $W_{\text{ph}%
} $. This excludes subspaces containing the Faddeev-Popov ghosts, the
temporal and longitudinal components of the gauge fields, etc. A quick way
to prove the preservation of the symmetries is by means of the techniques
recently developed in refs. \cite{AbsoPhys}. There, it was shown how to
dress the elementary fields to make them manifestly gauge invariant, without
altering the fundamental theory. Working with dressed fields, and recalling
that they reduce to the ordinary elementary fields at the level of on-shell
asymptotic states, it is evident that the operations involved in formula (%
\ref{ups}) are manifestly gauge invariant. As far as $\Omega _{\text{ph}}(A)$
is concerned, we can proceed as before, by comparison with \cite%
{diagrammarMio}, and recalling that: $a$) the projection to purely virtual
particles amounts to remove certain types of singularities everywhere from
Feynman diagrams; and $b$) different singularities do not talk to one
another.

Thus, once we assume that the physical space $W_{\text{ph}}$ is invariant
(e.g., it is made of physically observable particles), the projection $%
V\rightarrow V_{\text{ph}}(A)$ is invariant, and the preservation of
symmetries is always guaranteed, including the cancellation of anomalies to
all orders by means of the Adler-Bardeen theorem \cite{AdlerBardeen}.

The results of this paper extend to the off-shell amplitudes of gauge
invariant fields defined in \cite{AbsoPhys}. We also recall that a
particular projection is the one that gets rid of the Faddeev-Popov ghosts,
as well as the temporal and longitudinal components of the gauge fields, in
gauge theories. Applying the map $V\rightarrow V_{\text{ph}}(A)$ to that
case, we recover the proof of unitarity in gauge theories given in ref. \cite%
{LWScatteringLambda}.

\medskip

\textbf{Renormalizability}

The renormalizability of the projected amplitudes is manifest, whenever the
unprojected amplitudes are renormalizable. Indeed, formula (\ref{ups}) tells
us the reduced amplitude $V_{\Omega }$ is equal to the usual amplitude $V$
plus terms that involve one or more cuts. A single cut is sufficient to kill
the overall divergence of a diagram, since the delta function due to the cut
restricts the integration domain of the overall integral to a compact
subset. The subdivergences are automatically taken care of as usual. The $%
\Omega $ corrections are compatible with renormalizability as long as they
do not affect the zeroth levels of the threshold decomposition, as we have
required.

\medskip

Finally, we remark that the difference $V_{\Omega }-V$ between the projected
amplitude and the time-ordered one vanishes identically when the incoming
energy $E_{\text{in}}$ is smaller than the mass of the lightest purely
virtual particle. Indeed, the diagrams of $V_{\Omega }-V$ contain at least
one cut leg of type $\chi $, so they can be nontrivial only if $E_{\text{in}}
$ exceeds the $\chi $ mass. In particular, $V_{\Omega }-V$ vanishes in the
Euclidean region, and the renormalization of the projected theory coincides
with the one of the parent Euclidean theory.

\section{Conclusions}

\label{conclusions}\setcounter{equation}{0}

We have formulated a new quantization principle for quantum field theory,
based on a special type of non time-ordered product, and shown that it gives
the theories of physical and purely virtual particles.

The diagrams of the physical amplitude $V_{\text{ph}}=iT_{\text{ph}}$ are
built by means of the usual vertices and propagators, plus non time-ordered
propagators. The instructions to assemble the diagrams are encoded in a
formula that maps the standard amplitude $V=iT$ into $V_{\text{ph}}$. If $V$
obeys the unitarity or pseudounitarity equation, the most general reduced
amplitude $V_{\Omega }$ that obeys the unitarity equation depends on an
arbitrary anti-Hermitian matrix $\Omega $. A special $\Omega $, called $%
\Omega _{\text{ph}}(A)$ in section \ref{Symmetries}, is determined by
requiring that the projection of a product diagram is equal to the product
of the projected factors, and that the factorization survives basic
operations of ascent and descent through the diagrams. The idea is that the
time ordering should be violated in a sort of \textquotedblleft
minimum\textquotedblright\ way, inside prime diagrams. The amplitude $V_{%
\text{ph}}(A)=V_{\Omega _{\text{ph}}(A)}$ coincides with the amplitude $V_{%
\text{ph}}$ of a theory of physical and purely virtual particles, as given
in ref. \cite{diagrammarMio}.

We have worked out a number of techniques to relate different diagrams.
Besides descending from bigger to smaller diagrams, it is also possible to
ascend in a unique way from smaller to bigger diagrams, derive their $\Omega 
$ corrections, and match the threshold decompositions level by level. We
have illustrated these properties in various examples. At one loop, we have
considered the ascending chain bubble $\rightarrow $ triangle $\rightarrow $
box $\rightarrow $ pentagon $\rightarrow $ hexagon. At two loops, we have
focused on the first nontrivial arrangement, which is the box with diagonal.
In all the cases we have considered, nontrivial $\Omega $ corrections are
present when the diagram factorizes, and when it factorizes under the
contractions of some internal legs. Moreover, it is always possible to
ascend through the threshold decompositions in a unique way. We conjecture
that these are general properties of the physical amplitude $V_{\text{ph}}$.

Purely virtual particles provide the most elegant way to break the crystal
glass of time ordering. To give the reader an idea of how inelegant the most
general solution (\ref{ups}) is, consider that, when $\Omega $ is generic
(including $\Omega =0$), a diagram with non-amputated external legs is not
straightforwardly related to the same diagram with amputated external legs,
and has to be calculated anew. Moreover, the usual definitions of generating
functionals of connected and irreducible Green functions do not apply. Only
for $\Omega =\Omega _{\text{ph}}(A)$, we have that, if $Z(J)$ denotes the
generating functional all the correlation functions, its logarithm $%
W(J)=-i\ln Z(J)$ is the generating functional of the connected ones, and the 
$W$ Legendre transform $\Gamma (\Phi )=W(J)-\int \Phi J$, $\Phi =\delta
W/\delta J$, is the generating functional of the amputated, one-particle
irreducible ones.

Yet, we cannot exclude that unforeseen physical principles might one day
point to one of the many alternative options. In the absence of experimental
data, the only thing we can do is single out the options that have
remarkable formal and diagrammatic properties. In this spirit, it may be
also worth to search for alternative $\Omega $ corrections, which may break
the time ordering in non minimal ways, but have other interesting
properties. In any case, the right solution chosen by nature must lie
somewhere in formula (\ref{ups}), determined by the $\Omega $ that fits the
physics.

We conclude with a brief summary of the formulations of purely virtual
particles worked out so far. First, a nonanalytic Wick rotation was
introduced in refs. \cite{Piva}, as a way to get rid of ghosts with complex
masses, and reformulate the Lee-Wick models \cite{leewick}. Its key
ingredient is the average continuation around the branch cuts of amplitudes.
It was soon realized that the procedure was actually a way to formulate
models of new types, rather different from the original Lee-Wick idea (see 
\cite{LWfakeons} for a detailed comparison), and could be extended to remove
ghosts with real masses (as well as physical particles), to give sense of
quantum gravity as a power counting renormalizable theory, like the standard
model \cite{LWgrav}. The proof of unitarity to all orders in this approach
was given in ref. \cite{LWfakeons}\footnote{%
For Lee-Wick approaches to quantum gravity, we address the reader to refs. 
\cite{tomboulis}. Among other approaches to the problem of removing ghosts
in quantum field theory, we point out \cite{berends}.}.

The second, equivalent formulation of purely virtual particles was
introduced by means of the diagrammatic threshold decomposition of \cite%
{diagrammarMio}, and the spectral optical identities derived from it. The
third formulation, equivalent to the other two, is the one of the present
paper, based on the minimally non time-ordered product. With respect to \cite%
{LWfakeons}, the gain offered by the two new formulations is considerable,
not only for the clarity of the proofs to all orders, but also from the
practical point of view. Indeed, the new formulations offer several ways to
make calculations with not much more effort than computing Feynman diagrams
(check \cite{PivaMelis}).

We end by mentioning some perspectives for the future. One goal is to
develop the operatorial/Hamiltonian approach to purely virtual particles,
and maybe study their quantum mechanics, where the evolution operator is no
longer the usual time-ordered exponential, but follows from the minimally
non time-ordered product. Another challenging objective is to pursue the
off-shell formulation of transition amplitudes, by combining the approach of
this paper with the results recently obtained in refs. \cite{AbsoPhys},
where it was shown how to define off-shell physical amplitudes of colored
states in QCD\ and point-dependent observables in quantum gravity,
consistently with the (off-shell version of the) optical theorem.

All in all, we think that we have achieved a satisfactory understanding of
the nature of purely virtual particles, and revealed the main mysteries
behind them, in view of the challenges just mentioned.

\vskip 1truecm

\noindent {\textbf{\huge Appendices}} \renewcommand{\thesection}{%
\Alph{section}} \renewcommand{\theequation}{\thesection.\arabic{equation}} %
\setcounter{section}{0}

\vskip .5truecm

\section{From diagrams to scattering matrix, and back}

\label{diagrammar20}\setcounter{equation}{0}

Although the matrix $V=iT$ is a collection of diagrams, in the paper we have
been able to concentrate on single diagrams, and isolate the identities
satisfied by them. In this appendix we show how to switch from $T$ to single
diagrams, with a different particle in each internal leg (and vice versa),
with no loss and no gain of information.

For definiteness, we start from a Lagrangian $\mathcal{L}(\varphi )$ that
depends on a single field $\varphi $. We separate the kinetic (i.e.,
quadratic) part $\mathcal{L}_{\text{kin}}(\varphi ,m)$, where $m$ denotes
the mass of $\varphi $, from the interaction part $\mathcal{L}_{\text{int}%
}(\varphi ,\lambda )$ (made by anything that is not quadratic in $\varphi $,
including the linear terms, if present), where $\lambda $ denotes the
couplings: 
\begin{equation}
\mathcal{L}(\varphi )=\mathcal{L}_{\text{kin}}(\varphi ,m)+\mathcal{L}_{%
\text{int}}(\varphi ,\lambda ).  \label{L}
\end{equation}%
Then we use a Pauli-Villars trick \cite{PV} to introduce many fields $%
\varphi _{i}$ without changing the diagrams. Specifically, we replace the $%
\varphi $ kinetic part with the sum of the $\varphi _{i}$ kinetic parts,
having the same mass. Moreover, we replace $\varphi $ with the sum of all
the $\varphi _{i}$ in the interaction part:%
\begin{equation}
\mathcal{L}^{\prime }(\varphi )=N\sum_{i=1}^{N}\mathcal{L}_{\text{kin}%
}(\varphi _{i},m)+\mathcal{L}_{\text{int}}\left( \phi ,\lambda \right)
,\qquad \qquad \phi =\sum_{i=1}^{N}\varphi _{i}.  \label{Lp}
\end{equation}%
The diagrams $G_{E}$ with $E$ external legs, generated by this Lagrangian,
coincide with those generated by (\ref{L}), multiplied by $N^{-E/2}$ and an
appropriate combinatorial factor. To prove this, it is sufficient to note
that each internal leg carries the $\phi $ propagator, which is the sum of
the $\varphi _{i}$ propagators, which in turn is equal to the $\varphi $
propagator. We can choose the external legs $\varphi _{i}$ we want, and the
diagram $G_{E}$ is always the same, apart from the factors in front.

\medskip

At this point, we give a different mass $m_{i}$ to each field $\varphi _{i}$%
, and a different coupling $\lambda _{I}$ to each vertex obtained by
expanding the interaction part, where the subscript $I$ refers to the
various possibilities we have. We obtain%
\begin{equation*}
\mathcal{L}_{\text{ext}}(\varphi )=N\sum_{i=1}^{N}\mathcal{L}_{\text{kin}%
}(\varphi _{i},m_{i})+\mathcal{\tilde{L}}_{\text{int}}\left( \varphi
_{i},\lambda _{I}\right) ,
\end{equation*}%
for a certain, new interaction Lagrangian $\mathcal{\tilde{L}}_{\text{int}}$%
. We know that we retrieve the diagrams $G_{E}$ of the starting theory when
we set all the masses $m_{i}$ equal to $m$, and all the couplings $\lambda
_{I}$ equal to the appropriate values $\lambda $.

\medskip

Now we show that the extended Lagrangian $\mathcal{L}_{\text{ext}}(\varphi )$
allows us to isolate the diagrams as needed. Let $V=iT$ and $V_{\text{ext}%
}=iT_{\text{ext}}$ denote the usual (unprojected) amplitudes, associated
with $\mathcal{L}(\varphi )$ and $\mathcal{L}_{\text{ext}}(\varphi )$,
respectively. Let $G$ and $G_{\text{ext}}$ denote diagrams contributing to
them. For any $G$ there exists a generalization $G_{\text{ext}}$ with the
same topology as $G$, where each internal leg propagates a different field $%
\varphi _{i}$. It is sufficient to take $N$ sufficiently large, to have a
sufficient number of different fields $\varphi _{i}$, and differentiate $V_{%
\text{ext}}$ with respect to suitable couplings $\lambda _{I}$. After that,
we set $\lambda _{I}=0$ for every $I$.

\medskip

After building the diagram $G_{\text{ext}}$ with this method, we can study
the diagrammatic identities satisfied by it, and build the projections we
need, by differentiating formula (\ref{ups}) with respect to the appropriate
couplings. If the $\Omega $ corrections are determined as explained in the
paper, the right identities are obtained by differentiating (\ref{ups}) at $%
\Omega \neq 0$ as well.

\medskip

An advantage of $G_{\text{ext}}$ is that its combinatorics are trivial,
since there is only one Wick contraction for each field $\varphi _{i}$
participating in it. When we set the masses equal to one another, and
identify the couplings appropriately to go back to the original theory (\ref%
{L})-(\ref{Lp}), several diagrams give identical contributions and restore
the right combinatorics of $\mathcal{L}(\varphi )$.

\medskip

In the paper, we built diagrams with independent internal legs and no
external legs, replaced by external sources $K$. The sources can be replaced
by products of physical fields without affecting the results we have
obtained, as long as the external legs are amputated. When we need to
include propagators on the external legs, we can use the factorization
property, since we know that the right $\Omega $ gives the factorized
result. With a generic $\Omega $, instead, the factorization property does
not hold and we must redo the whole projection.

\section{Identities for product distributions}

\label{app2}\setcounter{equation}{0}

In this appendix we prove some identities for product distributions that we
have used in the paper. The first one is%
\begin{equation}
\mathcal{P}\int_{-\infty }^{+\infty }\frac{\mathrm{d}y}{2\pi }\frac{1}{y(x+y)%
}=\frac{\pi }{2}\delta (x),  \label{idainte}
\end{equation}%
where $x$ is real. It be proved by integrating the expression%
\begin{equation*}
\left[ \frac{1}{(x+i\epsilon )(y+i\epsilon )}-\frac{1}{(x+i\epsilon
)(x+y+2i\epsilon )}\right] +\frac{1}{(y+i\epsilon )(x+y+2i\epsilon )}=0,
\end{equation*}%
on $y$ from $-\infty $ to $+\infty $. Since the integral of the terms in
square brackets gives zero, we obtain%
\begin{equation}
\int_{-\infty }^{+\infty }\frac{\mathrm{d}y}{2\pi }\frac{1}{(y+i\epsilon
)(x+y+2i\epsilon )}=0.  \label{inte}
\end{equation}%
Formula (\ref{idainte}) then follows by using%
\begin{equation}
\frac{1}{x+i\epsilon }=\mathcal{P}\left( \frac{1}{x}\right) -i\pi \delta (x)
\label{deca}
\end{equation}%
twice inside the integral (\ref{inte}). Below, we reassure the reader that
it is correct to use the decomposition (\ref{deca}) in products.

More quickly, formula (\ref{idainte}) can be also proved from the identity%
\begin{equation}
\mathcal{P}\frac{1}{x}\left( \frac{1}{y}-\frac{1}{x+y}\right) -\mathcal{P}%
\frac{1}{y(x+y)}=-\pi ^{2}\delta (x)\delta (y),  \label{identa}
\end{equation}%
derived in ref. \cite{diagrammarMio}. The $y$ integral of the left term is
convergent and gives zero, so the rest gives (\ref{idainte}).

The second identity we need is%
\begin{equation}
\mathcal{P}\int_{-\infty }^{+\infty }\frac{\mathrm{d}z}{2\pi }\frac{1}{%
(x+z)(y+z)z}=\frac{\pi }{2}\mathcal{P}\left[ \frac{\delta (x)}{y}+\frac{%
\delta (y)}{x}-\frac{\delta (x-y)}{x}\right] .  \label{ide3}
\end{equation}%
We start from the formula%
\begin{equation*}
\mathcal{P}\left[ \frac{1}{xyz}-\frac{1}{x+y+z}\left( \frac{1}{xy}+\frac{1}{%
xz}+\frac{1}{yz}\right) \right] =0,
\end{equation*}%
which was derived again in ref. \cite{diagrammarMio}. First, we reflect $z$
to $-z$, then translate $x$ and $y$ by $z/2$, finally rescale $z$ by a
factor 2. Integrating on $z$, we get 
\begin{equation*}
0=\mathcal{P}\int_{-\infty }^{+\infty }\frac{\mathrm{d}z}{2\pi }\left[ \frac{%
1}{(x+z)(y+z)z}+\frac{1}{x+y}\left( \frac{2}{(x+z)(y+z)}-\frac{1}{(x+z)z}-%
\frac{1}{(y+z)z}\right) \right] .
\end{equation*}%
Every integral is separately convergent, so using (\ref{idainte}) we obtain (%
\ref{ide3}).

Now we show that it is correct to use the decomposition (\ref{deca}) in
products. Consider%
\begin{equation*}
\int \frac{\phi (x,y)}{(x+i\epsilon )(y+i\epsilon ^{\prime })},
\end{equation*}%
where $\phi (x,y)$ is a generic test function in two variables and the
integral is over the plane $xy$, with measure $\mathrm{d}x\mathrm{d}y/(2\pi
)^{2}$. Decompose $\phi (x,y)$ as the sum of%
\begin{equation*}
\phi _{\sigma \tau }(x,y)=\frac{1}{4}\left[ \phi (x,y)+\sigma \phi
(-x,y)+\tau \phi (x,-y)+\sigma \tau \phi (-x,-y)\right] ,
\end{equation*}%
according to the $x$ and $y$ parities, where $\sigma $ and $\tau $ can be $%
+1 $ or $-1$. Note that%
\begin{equation*}
\int \frac{\phi _{--}(x,y)}{(x+i\epsilon )(y+i\epsilon ^{\prime })}=\int 
\frac{\phi _{--}(x,y)}{xy}=\mathcal{P}\int \frac{\phi (x,y)}{xy},
\end{equation*}%
since $\phi _{--}(x,y)/(xy)$ is regular for $x\sim 0$ and $y\sim 0$.
Moreover, 
\begin{equation*}
\int \frac{\phi _{-+}(x,y)}{(x+i\epsilon )(y+i\epsilon ^{\prime })}=-i\int 
\frac{\epsilon ^{\prime }\phi _{-+}(x,y)}{(x+i\epsilon )(y^{2}+\epsilon
^{\prime 2})}=\mathcal{P}\int \frac{-i\pi \delta (y)}{x}\phi (x,y)
\end{equation*}%
and 
\begin{equation*}
\int \frac{\phi _{++}(x,y)}{(x+i\epsilon )(y+i\epsilon ^{\prime })}=-\int 
\frac{\epsilon \epsilon ^{\prime }\phi _{++}(x,y)}{(x^{2}+\epsilon
^{2})(y^{2}+\epsilon ^{\prime 2})}=-\int \pi ^{2}\delta (x)\delta (y)\phi
(x,y).
\end{equation*}

Thus,%
\begin{eqnarray*}
\int \frac{\phi (x,y)}{(x+i\epsilon )(y+i\epsilon ^{\prime })} &=&\int \frac{%
\phi _{++}(x,y)+\phi _{-+}(x,y)+\phi _{+-}(x,y)+\phi _{--}(x,y)}{%
(x+i\epsilon )(y+i\epsilon ^{\prime })} \\
&=&\int \left[ \mathcal{P}\frac{1}{x}-i\pi \delta (x)\right] \left[ \mathcal{%
P}\frac{1}{y}-i\pi \delta (y)\right] \phi (x,y).
\end{eqnarray*}

\end{document}